\newcommand*{\ATLASLATEXPATH}{latex/}
\pdfoutput=1
\documentclass[cernpreprint,texlive=2011,txfonts,UKenglish]{latex/atlasdoc}
\usepackage{\ATLASLATEXPATH atlaspackage}
\usepackage{\ATLASLATEXPATH atlasbiblatex}
\usepackage{\ATLASLATEXPATH atlasphysics}

\usepackage{multirow}

\graphicspath{{logos/}{figures/}}

\usepackage{DiphotonPublication-defs}

\AtlasTitle{Search for resonances in diphoton events at $\sqrt{s}$=13 TeV  with the ATLAS detector}

\author{The ATLAS Collaboration}

\AtlasRefCode{HIGG-2016-08}

\PreprintIdNumber{CERN-EP-2016-120}

\AtlasJournalRef{JHEP09(2016)001}
\AtlasDOI{10.1007/JHEP09(2016)001}

\AtlasAbstract{%
Searches for new resonances decaying into two photons in the ATLAS experiment at the
CERN Large Hadron Collider are described. The analysis is based on proton--proton collision data corresponding to an integrated
luminosity of 3.2~fb$^{-1}$ at $\sqrt{s} = 13$~TeV recorded in 2015. Two searches are performed,
one targeted at a spin-2 particle of mass larger than 500~GeV, using Randall--Sundrum graviton states as a benchmark
model, and one optimized for a spin-0 particle of mass larger than 200~GeV.
Varying both the mass and the decay width,
the most significant deviation from the
background-only hypothesis is observed at a diphoton invariant mass around 750 GeV with
local significances of 3.8 and 3.9 standard deviations in the searches optimized for
a spin-2 and spin-0 particle, respectively. The global significances are estimated to be
2.1 standard deviations for both analyses. The consistency between the data collected at 13~TeV and 8~TeV
is also evaluated. Limits on the production cross section times branching ratio to two photons for the two 
resonance types are reported.}

\hypersetup{pdftitle={ATLAS document},pdfauthor={The ATLAS Collaboration}}

\begin{document}

\maketitle

\section{Introduction}
\label{sec:intro}

New high-mass states decaying into two photons are predicted in many 
extensions of the Standard Model (SM).
The diphoton final state provides a clean experimental signature with excellent invariant mass resolution
and moderate backgrounds.
Searches for new high-mass resonances decaying into two photons are described,
using CERN Large Hadron Collider (LHC)~\cite{lhc} proton--proton ($pp$) collision data at $\sqrt{s}$=~13~TeV recorded in 2015 by the ATLAS detector.
The data correspond
to an integrated luminosity of 3.2~fb$^{-1}$.

The decay photons would have different kinematic properties depending on whether the hypothetical particle has spin-0 or spin-2.
These are exploited by applying two different selections, 
with looser kinematic selection requirements for the spin-2 resonance search. 
The photon identification criteria and the event pre-selection are common to both searches.

The search for a spin-2 $\gamma\gamma$ resonance uses the
Randall--Sundrum~(RS) model~\cite{Randall:1999ee} graviton as a benchmark. This entails a lightest Kaluza--Klein~\cite{ModernKK} spin-2
graviton excitation ($G^*$) with a dimensionless coupling \coupling , where $\mplRed = \mpl/\sqrt{8\pi}$ is the reduced Planck scale
and $k$ the curvature scale of the extra dimension. 
The lightest graviton excitation is expected to be a fairly narrow resonance for \coupling < 0.3~\cite{Davoudiasl:1999jd},
with the width given by $1.44 (\coupling)^2 m_{G^*}$, where $ m_{G^*}$ is the mass of the lightest graviton state.
For $\coupling = 0.1$, the natural width increases from $11$~GeV at $\mG = 800$~GeV to $30$~GeV at $\mG = 2200$~GeV.
For $\mG = 800$~GeV, the contributions of the natural width and of the experimental mass resolution 
to the width of the resonance are comparable.
The shape of the invariant mass distribution of the main background from the production of prompt photon pairs is estimated from theoretical computations, 
and the contribution from the reducible background of events where at least one jet is misidentified as a photon is added from data-driven estimates. 
This approach works well  up to the highest invariant masses where a small number of background events are expected.
The search is performed in the mass range above 500~GeV
and in the \coupling\ range 0.01 to 0.3, searching for an excess over the estimated background diphoton invariant mass distribution.
To model such an excess, the RS graviton resonance shape is convolved with the experimental resolution.

Spin-0 $\gamma\gamma$ resonances are predicted in
theories with an extended Higgs sector~\cite{Hill:1987ea,Veltman:1989vw,Binoth:1996au,Schabinger:2005ei,Patt:2006fw,Pruna:2013bma,Lee:1973iz}. 
The search for a spin-0 resonance uses a restricted
kinematic range for the photon selection, taking advantage of the isotropic distribution of the decay products in  the centre-of-mass frame of the new particle.
The background is estimated by fitting the diphoton invariant mass distribution to an analytical function,
searching for an excess modelled by a spin-0 resonance convolved with the
experimental resolution. 
The search is performed in the mass range 200--2000~GeV, where there are enough events to constrain  the background shape, and
for width values up to 10\% of the mass of the hypothesized particle.

Searches for diphoton resonances in LHC Run-1 data have been reported by the ATLAS and CMS collaborations~\cite{Aad:2012cy,Chatrchyan:2011fq,Aad:2015mna,Aad:2014ioa,Khachatryan:2015qba}.
A similar analysis was performed by the CMS collaboration using the 2015 LHC pp data~\cite{CMS-diphoton-2015}.

This paper is organized as follows. After a description of the ATLAS detector in Section~\ref{sec:detector}
and of the data and simulated event samples in Section~\ref{sec:samples}, 
the photon selection and energy measurements are presented in Section~\ref{sec:photon_selection}. 
In Sections~\ref{sec:event_selection} to~\ref{sec:stat} the event selection, the modelling of the signal and the estimation of the background 
as well as the statistical procedure to analyse the data are presented. The results are discussed in Section~\ref{sec:results}.
 
\section{ATLAS detector}
\label{sec:detector}

The ATLAS detector~\cite{PERF-2007-01} 
is a multi-purpose detector with a forward-backward symmetric
cylindrical geometry.\footnote{The ATLAS experiment uses a right-handed coordinate system with its origin at the
nominal interaction point (IP) in the centre of the detector and the $z$-axis along the beam
pipe. The $x$-axis points from the IP to the centre of the LHC ring, and the $y$-axis points
upward. Cylindrical coordinates $(r,\phi)$ are used in the transverse plane, $\phi$ being the
azimuthal angle around the beam pipe. The pseudorapidity is defined in terms of the polar angle
$\theta$ as $\eta=-\ln\tan(\theta/2)$.} At small radii, the inner detector (ID), immersed in a 2~T
magnetic field produced by a thin superconducting solenoid located in front of the calorimeter, 
is made up of fine-granularity pixel and microstrip detectors. These silicon-based detectors cover
the pseudorapidity range $|\eta|<2.5$. A gas-filled straw-tube transition radiation tracker (TRT)
complements the silicon tracker at larger radii and also provides electron identification capabilities based on
transition radiation. The electromagnetic (EM) calorimeter is a lead/liquid-argon sampling
calorimeter with accordion geometry.  The calorimeter is divided into a barrel section covering
$|\eta|<1.475$ and two end-cap sections covering $1.375<|\eta|<3.2$. For $|\eta|<2.5$ it is divided
into three layers in depth, which are finely segmented in $\eta$ and $\phi$. A thin presampler
layer, covering $|\eta|<1.8$, is used to correct for fluctuations in upstream energy losses.
Hadronic calorimetry in the region $|\eta|<1.7$ uses steel absorbers and scintillator tiles as the
active medium.  Liquid-argon calorimetry with copper absorbers is used in the hadronic end-cap
calorimeters, which cover the region $1.5<|\eta|<3.2$.  A forward calorimeter using copper or
tungsten absorbers with liquid argon completes the calorimeter coverage up to $|\eta|=4.9$.  The
muon spectrometer measures the deflection of muon tracks within $|\eta|<2.7$, using three
stations of precision drift tubes, with cathode strip chambers in the innermost layer for
$|\eta|>2.0$. 
The deflection is provided by a toroidal magnetic field from air-core superconducting magnets,
with an integral of approximately 3~T$\cdot$m and 6~T$\cdot$m 
in the central and end-cap regions, respectively. The muon spectrometer is instrumented with 
trigger chambers covering $|\eta|<2.4$.
Events are selected using a first-level trigger implemented in custom electronics, 
which reduces the event rate to a design value of 100~kHz using a subset of detector information. 
Software algorithms with access to the full detector information are then used in the high-level trigger to yield a 
recorded event rate of about 1~kHz~\cite{ATL-DAQ-PUB-2016-001}.

\section{Data and simulated event samples}
\label{sec:samples}

Data were collected by the ATLAS detector in 2015 using $pp$ collisions at a centre-of-mass energy of $\sqrt{s}$~=~13~TeV 
with a minimum bunch spacing of 25~ns, an average number of $pp$ interactions per bunch crossing of about 13, and a peak instantaneous luminosity of $5~\times 10^{33}$ cm$^{-2}$s$^{-1}$.
Events from $pp$ collisions were recorded using a diphoton trigger with transverse energy $\et = E \sin(\theta)$
thresholds of 35~GeV and 25~GeV
for the \et-ordered leading and subleading photon candidates, respectively. 
In the high-level trigger, clusters of energy in the EM calorimeter are reconstructed and required to satisfy loose criteria according to the properties of showers initiated by photons.
The trigger has a signal efficiency close to 99\% for events fulfilling the final event selections.
Only events taken in stable beam conditions, and in which the detector is fully operational,
are considered. 
After data-quality requirements, the data sample corresponds to an integrated luminosity of 3.2~fb$^{-1}$. 
The measurement of the integrated luminosity has an uncertainty of $\pm$5\%. 
It is derived, following a methodology similar to that detailed in Ref.~\cite{Aad:2013ucp},
from a preliminary calibration of the luminosity scale using van der Meer scans performed in August 2015.

Simulated Monte Carlo (MC) samples are used to optimize the search strategy and to study background sources.
Interference effects between signal and background processes are neglected.
Signal samples for the RS graviton model are generated using \textsc{Pythia8}~\cite{Pythia8}, version 8.186, 
with the NNPDF23LO~\cite{nnpdf23lo}  parton distribution functions (PDF) and the A14~\cite{ATL-PHYS-PUB-2014-021}
set of tuned parameters (tune) for the underlying event,
for different
choices of the graviton mass and the parameter \coupling\ , spanning a mass range from 500~GeV to 5000~GeV and
\coupling\ values from 0.01 to 0.3.
Only the lowest-mass RS graviton state is generated. 
Samples for any mass or \coupling\ value are obtained
by reweighting  an event sample generated with a uniform mass distribution
using the Breit--Wigner and parton luminosity terms.
The latter sample is obtained using the \textsc{Pythia8} configuration as described above with $m_{G^*}$ = 5~TeV and \coupling\ = 0.1, 
with an $m_{\gamma\gamma}$-dependent factor that modifies the production cross section to remove the effect of the Breit--Wigner term and the parton luminosity.
The validity of this procedure is verified using samples generated at discrete
values of $m_{G^*}$ and \coupling.  Because the graviton coupling increases with the energy of the decay products, 
the invariant mass distribution for a large-width
graviton signal exhibits a high-mass tail which is not present for a spin-0 particle with properties like those of a Higgs boson.

The signal in the spin-0 particle search is simulated as if it were a SM Higgs boson produced in $pp$ collisions
via gluon fusion and decaying into two photons. Other production processes are investigated to assess the impact
of the production mode on the signal modelling.
MC samples are produced for different hypotheses of the
spin-0 boson mass ($m_X$) in the range 200~GeV to 2000~GeV and of the decay width ($\Gamma_X$) up to 10\% of $m_X$.
For the narrow-width approximation (NWA), the width
of the particle is set to 4~MeV. Gluon fusion events are generated with \textsc{Powheg-box}~\cite{powhegbox,Bagnaschi:2011tu}, version 2, interfaced with
\textsc{Pythia8} for the underlying event, parton showering and hadronization. 
To model signals with large decay widths,
a function parameterizing the theoretical line-shape of the resonance is used~\cite{Aad:2014ioa,Khachatryan:2015qba}.
The \textsc{Powheg-box} implementation of a large-width spin-0 resonance with couplings like those of the Higgs boson 
in the SM is chosen for this
function. 
The line-shape is modelled with a Breit--Wigner distribution based on a running-width scheme, including the dependence 
of the cross section on the gluon--gluon parton luminosity.
In order to reduce the sensitivity to modelling effects from the off-shell region, the sample generation is restricted to the region
$m_X \pm 2 \Gamma_X$. 
The validity of this procedure is checked by comparing the result of this implementation with simulated samples generated with a 
large width in \textsc{Powheg-box}.
Events produced via vector-boson fusion
are generated using \textsc{Powheg-box}~\cite{Nason:2009ai}  interfaced with \textsc{Pythia8}. Associated production with a vector-boson or a $t\bar{t}$ pair
is generated with \textsc{Pythia8}. The \textsc{CT10}~\cite{CT10} PDF set is used for the samples generated with \textsc{Powheg-box}, while
\textsc{CTEQ6L1}~\cite{Pumplin:2002vw} is used for the samples generated with \textsc{Pythia8}. 
The underlying-event generation for the gluon fusion and vector-boson fusion samples
is based on the \textsc{Pythia8} AZNLO tune~\cite{AZNLO:2014}, while for the other samples, the A14 tune is used.

Events containing two prompt photons,
representing the largest irreducible background to the search, 
are simulated using the \textsc{Sherpa}~\cite{Gleisberg:2008ta} generator version 2.1.1. 
Matrix elements are calculated with up to two partons at leading order in QCD and merged with the \textsc{Sherpa} parton shower~\cite{Schumann:2007mg}
 using the ME+PS@LO prescription~\cite{Hoeche:2009rj}.
The gluon-induced box process is also included.
The \textsc{CT10} PDF set is used in conjunction with a dedicated parton-shower tune of  \textsc{Sherpa}.
Samples of the photon+jet reducible background component are also generated using  \textsc{Sherpa}.
For comparisons, \textsc{Pythia8} is also used to generate Standard Model diphoton production, based on the leading-order quark--antiquark $t$-channel 
annihilation diagram
and the gluon-induced box process, and photon+jet production.

The generated events are passed through a full detector simulation~\cite{Aad:2010ah} based on \textsc{Geant4}~\cite{Geant4}. 
Pile-up from additional $pp$ collisions  in the same and  neighbouring bunch crossings is simulated by overlaying each MC
event with a variable number of simulated inelastic $pp$ collisions 
generated using \textsc{Pythia8} with the A2 tune~\cite{ATL-PHYS-PUB-2012-003}. The MC events are weighted to reproduce
the distribution of the average number of interactions per bunch crossing observed in the data.

\section{Photon selection}
\label{sec:photon_selection}

Photon and electron candidates are reconstructed from clusters of energy deposited in the electromagnetic
calorimeter.  Candidates without a matching track or reconstructed conversion vertex in the inner detector are
classified as unconverted photons.  Those with a matching reconstructed conversion
vertex or a matching track, consistent with originating from a photon conversion, are classified as
converted photons.  Those matched to a track consistent with originating from an
electron produced in the beam interaction region are kept as electrons.

Only photon candidates with $|\eta| < 2.37$ are considered, not including
the transition region $1.37<|\eta|<1.52$ between the barrel and end-cap calorimeters.  The calorimeter
granularity in the excluded transition region is reduced, and the presence of significant additional inactive
material degrades the photon identification capabilities and energy resolution.

Photon identification is based primarily on shower shapes in the calorimeter~\cite{ATLAS-PERF-2013-04},
with the selection criteria re-optimized for the conditions expected for the 2015 data.
An initial loose selection is derived using only the information from the hadronic calorimeter and the lateral shower
shape in the second layer of the electromagnetic calorimeter, which contains most of the energy. 
The final tight selection applies tighter criteria to these variables, different for converted and unconverted photon
candidates. It also places requirements on the shower shape in the finely segmented first calorimeter layer 
to ensure the compatibility of the measured
shower profile with that originating from a single photon impacting the calorimeter.
When applying the photon identification criteria to simulated events, the shower shapes are corrected for small
differences in their average values between data and simulation.
The efficiency of the photon identification increases with \et\ from 85\% at 50~GeV to 95\% at 200~GeV.
For  \et\ $>$~50 GeV, the uncertainty in the photon identification efficiency varies between $\pm$1\% and $\pm$5\% depending
on $\eta$ and \et. 
This uncertainty is estimated from the effect of differences between shower-shape variable distributions in data and simulation.
From the studies done in  Ref.~\cite{ATLAS-PERF-2013-04},
this procedure is found to provide a conservative estimate of the uncertainties.

To further reject the background from jets misidentified as photons, the photon candidates are required
to be isolated using both calorimeter and tracking detector information.  The
calorimeter isolation variable, $\et^{\rm iso}$, is defined as the sum of the \et\ of energy clusters deposited in a cone
of size $\Delta R = \sqrt{(\Delta\eta)^2 + (\Delta\phi)^2} = 0.4$ around
the photon candidate, excluding an area of size
$\Delta \eta \times \Delta \phi = 0.125\times0.175$ centred on the
photon cluster; the expected photon energy deposit outside the
excluded area is subtracted. The pile-up and underlying-event contribution to the
calorimeter isolation variable is subtracted from the isolation energy
event-by-event~\cite{STDM-2010-08,Cacciari:2008gn,Cacciari2010}.  
The selection requirement on the calorimeter isolation variable is defined by $\et^{\mathrm{iso}} < 0.022 \et + 2.45$~GeV,
where $E_{\mathrm{T}}$ is the transverse energy of the photon candidate.
The track isolation variable ($\pt^{\mathrm{ iso}}$) is
defined as the scalar sum of the transverse momenta of the tracks in a 
cone of $\Delta R = 0.2$ around the photon candidate. The tracks are required to have $\pt > 1.0$~GeV and
to be consistent with originating from the diphoton primary vertex, defined in Section~\ref{sec:pv_selection}.
For converted photons, the one or two tracks associated with the photon conversion are
excluded from the $\pt^{\mathrm{iso}}$ computation.
The requirement applied for the track isolation variable is $\pt^{\mathrm{iso}} < 0.05  \et$.

The efficiency of the isolation requirements is studied using several data control samples.  Electrons from $Z$-boson decays
are used to validate the isolation variables up to \et = 100~GeV. Inclusive photon samples are used to check the 
efficiency of the isolation requirement 
in a wide \et range  from 50~GeV up to 1000~GeV.
Small differences between data and simulation in the average value of the calorimeter
isolation variable are observed as a function of \et and $\eta$ of the photon candidates.
The size of this difference is used as a systematic uncertainty. 
The efficiency of the combined isolation requirement for photons fulfilling tight identification selection in signal MC
samples is 90\% to 96\% in the \et\ range 100~GeV to 500~GeV, with an uncertainty between 1\% and 2\%. 
The isolation requirement reduces the 
rate at which jet are misidentified as photons by about one order of magnitude.

The measurement of the electron or photon energy is based on the energy collected in calorimeter
cells in an area of size $\Delta \eta \times \Delta \phi =  0.075\times0.175$  in the barrel
and $0.125\times0.125$ in the end-caps.  
A multivariate regression algorithm~\cite{Aad:2014nim} to
calibrate electron and photon energy measurements was developed and optimized on
simulated events. Corrections are made for the energy deposited in front of the calorimeter 
and outside the cluster, 
as well as to account for the variation of the energy response as a
function of the impact point on the calorimeter. The inputs to the energy calibration algorithm are
the measured energy per calorimeter layer, including the presampler, the $\eta$ of the cluster and the
local position of the shower within the second-layer cell corresponding to the cluster centroid. In
addition, for converted photons, the track transverse momenta and the conversion radius are used 
to further improve the energy resolution, especially at low
energy.  The  calibration of the layer energies in the calorimeter is based on the measurement
performed with 2012 data at $\sqrt{s} = 8$~TeV~\cite{Aad:2014nim}. The overall energy scale in data and
the difference in the constant term of the energy resolution 
between data and simulation are estimated with a sample of
$Z$-boson decays to electrons recorded in 2012 and reprocessed using the same conditions as used for the
2015 data taking and event processing. 
At \et\ values larger than $\approx$~200~GeV, the energy resolution is dominated by the constant term 
of the calorimeter energy resolution, which amounts to 
0.6\%--1.5\% depending on $\eta$.
The energy scale and resolution corrections are checked using $Z$-boson decays to electrons recorded
in the 2015 dataset. Uncertainties in the measurements performed with this sample are estimated following
a procedure similar to that discussed in Ref.~\cite{Aad:2014nim}. The difference between the values measured with the
2015 data and those predicted from the reprocessed 2012 data is also taken into account in the uncertainties.
The uncertainty in the photon energy scale at high \et is typically $\pm$(0.5--2.0)\% depending on $\eta$, 
and the relative uncertainty
in the photon energy resolution for \et\ =~300~GeV  is $\pm$(30--45)\% depending on $\eta$.
Additional uncertainties related to the extrapolation of the energy scale to photons of very high energies,
in addition to those described in Ref.~\cite{Aad:2014nim}, were considered and found to be small.
In particular, detailed checks of the validity of the calibration for the lowest gain range of the electronic readout~\cite{1748-0221-5-09-P09003} of the
electromagnetic calorimeter, 
which is used in the \et range above 350~GeV in the central part of the electromagnetic barrel calorimeter, were performed,
including checks with high-\et electrons from $Z$-boson decays.
These checks show that the relative calibration of the low-gain readout with respect to the other gains is better than $\pm$1\%.

\section{Event selection and sample composition}
\label{sec:event_selection}

Starting from the triggered events, two photon candidates fulfilling the tight identification criteria are required, with \et above 40~GeV and 30~GeV,
respectively.
The primary vertex corresponding to the $pp$ collision that produced the diphoton candidate is identified.
In addition, the calorimeter- and track-based isolation requirements
are applied to further reduce the background from jets misidentified as photons, thus increasing the
expected sensitivity of the analyses.
Different additional selections are then applied, separately in the spin-2 and spin-0 resonance searches.

\subsection{Primary vertex selection}
\label{sec:pv_selection}

The diphoton mass reconstruction requires the reconstructed primary vertex corresponding to the $pp$ collision that produced the
diphoton candidate.
The correct identification 
of the tracks originating from this $pp$ collision is also necessary to avoid
pile-up contributions to the track isolation.  To keep the contribution of the opening angle
resolution to the mass resolution smaller than the contribution of the energy resolution, a
position resolution for the primary vertex of about 15 mm in the $z$-direction is required.
Better resolution is needed to correctly match tracks to the $pp$ collision vertex of the
diphoton candidate.  The directions of both photon candidates are measured using the longitudinal
and transverse segmentation of the electromagnetic calorimeter, with a resolution of about 60~mrad/$\sqrt{E}$, where
$E$ is the photon energy in GeV. An estimate of the $z$-position of the diphoton primary vertex 
is obtained by combining the average beam-spot position with this `photon pointing'. It may be enhanced
using the tracks from photon conversions with conversion radii before
or in the volume of the silicon detectors. This estimate gives a resolution of about
15~mm in the $z$-direction. 
In order to select the correct primary vertex for the diphoton event,
a neural-network discriminant, similar to the one used in Ref.~\cite{HIGG-2013-12},
is constructed using both the  $z$-position of the diphoton primary vertex estimated
by the photon pointing including its uncertainty and
additional information from the tracks associated with each reconstructed primary vertex.
After applying this procedure, the contribution of the opening angle resolution
to the mass resolution is  negligible. The efficiency to reconstruct the correct
primary vertex within $\pm$0.3~mm is about 88\%.

\subsection{Event selection}

In the selection used to search for a spin-2 resonance, the transverse energy of each photon is required to satisfy
\et~$>$~55~GeV. 
With this selection, 5066 diphoton events with a diphoton invariant mass $m_{\gamma\gamma} >$~200~GeV are selected in the data.

The search for a spin-0 resonance applies tighter selections which
were optimized on simulated background and signal samples. 
Given the isotropic distribution of the decay,
the average transverse energy of the two photons is expected to be higher than that of photons from background processes
at the same invariant mass.
For a given value of  $m_{\gamma\gamma}$,
the transverse energy is required to be $\et > 0.4 m_{\gamma\gamma}$ for the photon with the highest \et\
and $\et > 0.3 m_{\gamma\gamma}$ for the photon with the second-highest $\et$.
This selection  improves the expected
sensitivity by more than 20\% for masses larger than 600~GeV compared to the initial requirement.
With these requirements, 7391 (2878)~events are selected in the data with $m_{\gamma\gamma} >$~150~GeV ($>$~200~GeV).
The highest invariant mass value observed in the data is 1933 GeV (1606 GeV) for the spin-2 (spin-0) search selection.

\subsection{Sample composition}
\label{sec:sample_composition}

The selected samples mainly consist of events 
from diphoton production, followed by photon+jet production,
 with one jet misidentified as a photon, and dijet
production with two jets misidentified as photons. Background sources from Drell--Yan, $W\gamma$ or $Z\gamma$ production, 
with either one or two
isolated electrons misidentified as photons, are negligible.
 A quantitative understanding of the sample composition is required for the background estimate in the spin-2 resonance
search. It is also used in the studies for the choice of background function in the spin-0 resonance search.

Two methods based on control regions built from events failing the isolation requirement and/or some of the tight photon identification
requirements are used to estimate the relative contribution of the
various sources of background directly from data.  
To avoid significant correlation with the isolation variable, 
only some of the tight photon identification requirements using the first layer of the calorimeter are inverted

In the first method~\cite{STDM-2012-05}, denoted the 2$\times$2 sidebands method, four regions for each photon candidate
are constructed, one region corresponding to the signal selection and the others to candidates failing the isolation requirement only,
failing part of the tight identification requirement only or failing both. For diphoton candidates, 16 control regions are thus obtained.
The inputs to the method are the numbers of events in the 16 regions and the signal efficiencies of the tight identification and isolation requirements. 
The correlation between these two requirements is assumed to be negligible for background events. 
The method allows the simultaneous extraction of the numbers of genuine diphoton events, photon+jet, jet+photon and dijet background events, 
and of the efficiencies of the tight identification and isolation requirements for photon candidates from misidentified jets.
Photon+jet events correspond to the cases where the sub-leading photon candidate
in \et\ is a jet misidentified as a photon, 
and vice versa for jet+photon events.

The second method~\cite{STDM-2011-05}, denoted the matrix method,
classifies the diphoton candidates passing tight identification requirements into four categories depending on whether both, 
only the leading,
only the sub-leading or none of the photons pass the isolation cut. 
The numbers of observed events in data in these categories are related to the
numbers of genuine diphoton, photon+jet, jet+photon and dijet events through isolation efficiencies for signal and background. 
The efficiency for background is estimated in control regions of the data, using events failing a 
subset of the tight identification requirements.
Events satisfying the tight identification are used to estimate the efficiency for genuine photons, 
after subtracting the background component, whose amount is estimated by comparing the number of events 
passing with the number failing a subset of the tight identification requirements, 
in control region of the data with large track isolation value, $\pt^{\mathrm{iso}}>0.05\et+10$~GeV.
Once these efficiencies are known, the sample composition can be extracted by the inversion of a 4$\times$4 matrix.
 
\begin{figure}[!]
\begin{center}
\subfloat[]{\includegraphics[width=.5\textwidth]{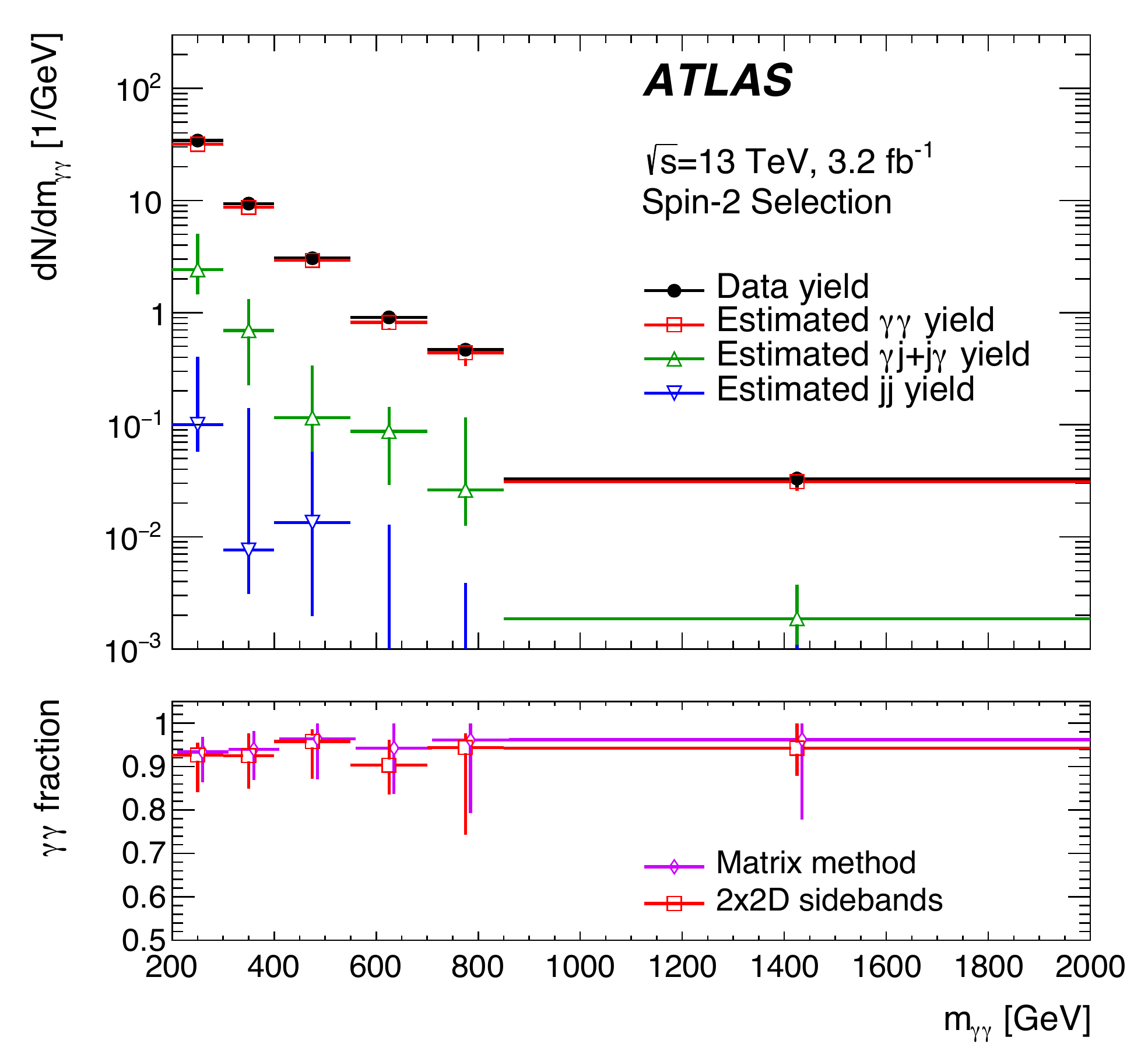}}
\subfloat[]{\includegraphics[width=.5\textwidth]{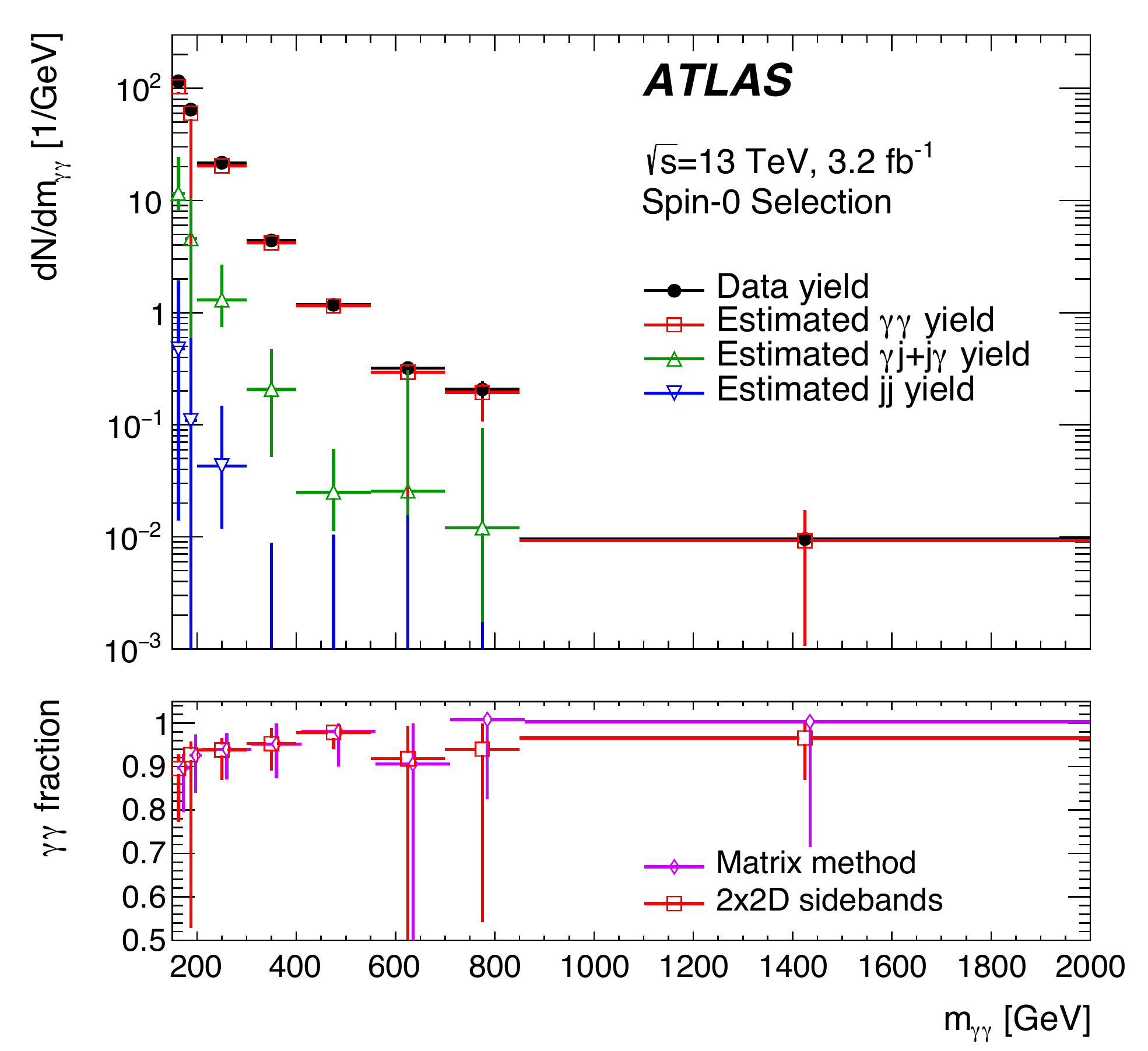}}
\end{center}
\caption{The diphoton invariant mass distributions (upper panels) of the data for the spin-2 
and spin-0 selections and their decomposition into contributions 
from genuine diphoton, photon+jet plus jet+photon and jet+jet events, determined as described in the text. The bottom panels show the purity of diphoton events as determined from the two methods.  The total uncertainties are shown, including statistical and systematic components.
}
\label{fig:bkg_composition}
\end{figure}

Both methods can be applied over the full selected kinematic range, or in bins of $m_{\gamma\gamma}$, thus providing inclusive 
as well as differential yields.
Figure~\ref{fig:bkg_composition} shows the decomposition of the selected data sample into the contributions from diphoton, 
photon+jet or jet+photon,
and dijet events
for both selections and the corresponding purities, defined as the ratio of diphoton events to the total number of events in the sample.
The purity is $(94^{+3}_{-7})$\% for the spin-2 selection and $(93^{+3}_{-8})$\% for the spin-0 selection. 
Uncertainties in these purity
estimates originate from the statistical uncertainty in the data sample, 
the definition of the control region failing the tight identification requirement, the modelling of the
isolation distribution and
possible correlations between the isolation variable and the inverted identification criteria. The two methods give consistent
results within their uncertainties. The estimate of these uncertainties is sensitive to the small number of events in some of the control regions.

\subsection{Signal acceptance and efficiency}
\label{sec:signalYield}

The expected signal yield can be expressed as the product of three terms: 
the production cross section times branching ratio to two photons, the acceptance (A) of the kinematic requirements, 
and the reconstruction and identification efficiency (C).
The acceptance
is expressed as the fraction of decays satisfying the fiducial acceptance at the generator level. 
The factor $C$ is defined as the ratio of the number of events 
fulfilling all the selections placed on reconstructed quantities to the number of events in the fiducial acceptance.
The fiducial acceptance closely follows
the selection criteria applied to the reconstructed data:  $|\eta_{\gamma}|<2.37$, $\et > 55$~GeV for the spin-2 resonance search
selection and $\et >0.4m_{\gamma\gamma}$ (leading $\gamma$), $\et >0.3m_{\gamma\gamma}$ (sub-leading $\gamma$) for the spin-0 resonance search.
An isolation requirement is applied in the fiducial acceptance definition using all particles with lifetime greater than 10~ps at the generator
level in a cone of $\Delta R = 0.4$ around the photon direction $\et^{\mathrm{iso}}<0.05 \et^{\gamma}+6$~GeV. The value
of the isolation requirement applied at the particle level is adjusted to reproduce the selection applied at the reconstruction level.

For the spin-2 resonance search, the acceptance for the benchmark RS graviton model varies from 66\% at a mass of 500 GeV,
to 91\% at a mass of 5000~GeV.
The factor $C$ is almost constant at 68\% in this mass range.
The value of $A \times C$ for the selection thus ranges from 
45\% to 61\% for masses between 500~GeV and 5000~GeV with a small dependence on the width. 

For the spin-0 resonance search, 
$A$ ranges from 52\% to 62\% in the mass range from 200~GeV to 700~GeV for a particle similar to a Higgs boson produced
by gluon fusion and is almost constant above 700~GeV.
The gluon fusion production mode is used to compute the value of $C$, which ranges from 65\% for a particle of mass 200~GeV to
71\% at 700~GeV and is almost constant above 700~GeV.
For the generator-level fiducial acceptance definition, the generated diphoton invariant mass is required to be
within $\pm2 \Gamma$ of the resonance mass.
Different
production modes (vector-boson fusion, associated production with a $W$ or $Z$ boson or with a $t\bar{t}$ pair) yield differences
in $C$ values of at most $\pm$3\%, which is taken as an uncertainty.
In the case of a decay width larger than the detector resolution,
the correction factor $C$ varies by up to $\pm$5\% depending on the assumed decay width.
This variation is taken as an additional uncertainty.

Experimental uncertainties in $C$ arise from uncertainties in the photon identification efficiency ($\pm$3\% to $\pm$2\% depending on the assumed mass
and on the selection), the photon isolation efficiency ($\pm$4\% to $\pm$1\% depending on the assumed mass and on the selection),
and the trigger efficiency  ($\pm$0.6\%). Uncertainties in $C$ related to the photon energy scale and resolution have a negligible impact
on the uncertainty in the expected signal yield.

\section{Signal modelling}
\label{sec:signal_modelling}

The invariant mass distribution of the diphoton pair for the signal is expected to peak near the
assumed mass of the new particle, with a spread given by the convolution of its intrinsic decay width
with the experimental resolution.
For both searches, the experimental resolution of the invariant mass is modelled with a double-sided
Crystal Ball (DSCB) function. Interference effects between signal and background are ignored.

The DSCB function is defined as:
 \begin{equation}
    N\cdot 
    \begin{cases}
    \mathrm{e}^{-0.5 t^{2}} & \mbox{if $-\alpha_{\mathrm{low}} \leq t \leq \alpha_{\mathrm{high}}$}\\
    \mathrm{e}^{-0.5\alpha_{\mathrm{low}}^{2} } \left[\frac{\alpha_{\mathrm{low}}}{n_{\mathrm{low}}} \left(\frac{n_{\mathrm{low}}}{\alpha_{\mathrm{low}}} - \alpha_{\mathrm{low}} -t \right)\right]^{-n_{\mathrm{low}}}  & \mbox {if $t < -\alpha_{\mathrm{low}} $}\\
    \mathrm{e}^{-0.5\alpha_{\mathrm{high}}^{2} } \left[\frac{\alpha_{\mathrm{high}}}{n_{\mathrm{high}}} \left(\frac{n_{\mathrm{high}}}{\alpha_{\mathrm{high}}} - \alpha_{\mathrm{high}}  + t \right)\right]^{-n_{\mathrm{high}}}  & \mbox {if $t > \alpha_{\mathrm{high}}$},\\
    \end{cases}
    \label{eq:DSCB}
 \end{equation}
 where $t = (m_{\gamma\gamma} - \mu_{\mathrm{CB}})/\sigma_{\mathrm{CB}}$,  $N$ is a normalization parameter, 
$\mu_{\mathrm{CB}}$ is the peak of the Gaussian distribution, 
$\sigma_{\mathrm{CB}}$ represents the width of the Gaussian part of the function,
$\alpha_{\mathrm{low}}$ ($\alpha_{\mathrm{high}}$) parameterizes the mass value where the 
distribution of the invariant mass
becomes a power-law function on the low-mass (high-mass) side, with
$n_{\mathrm{low}}$ ($n_{\mathrm{high}}$) the exponent of this function.
For samples with small decay width, the width of the DSCB Gaussian core $\sigma_{\mathrm{CB}}$ parameterizes 
the entire effect of the experimental 
invariant mass resolution.

The diphoton invariant mass resolution for a narrow resonance, as measured by the $\sigma_{\mathrm{CB}}$ parameter, varies from about 2~GeV
at a mass of 200~GeV to about 13~GeV at a mass of 2000~GeV. The relative uncertainty in the signal mass resolution is mostly driven by the
uncertainty in the constant term of the energy resolution, which is the dominant contribution at high
energy and varies from $^{+30}_{-20}$\% to $^{+60}_{-40}$\% as a function of the mass, in the range from 200~GeV to 1000~GeV and stays
almost constant above 1000~GeV.

For the spin-2 resonance search, the signal mass distribution for any value of the mass and \coupling\
is obtained by a convolution of the intrinsic detector resolution, modelled by a DSCB function,
with the predicted distribution of the mass line-shape at generator level, as discussed in  Section~\ref{sec:samples}.
The parameters of the DSCB function are determined from RS graviton signal samples of various masses with
$\coupling  = 0.01$, corresponding to a width of $1.14 \cdot 10^{-4}  m_{G^*}$, which is negligible compared to the detector resolution. The convolution approach takes
into account the high-mass tail predicted for the benchmark RS graviton model for large coupling values. It is validated by comparing
the predicted mass distribution to the one derived in fully simulated samples with different \coupling\ values and good agreement is found.

When considering spin-0 resonances with larger natural widths, simulated as discussed in Section~\ref{sec:samples}, 
the reconstructed line-shapes for a spin-0 signal are well described by DSCB functions. 
The function effectively parameterizes the combined effects of the theoretical line-shape and the detector response.
The parameters of the DSCB fit function are then expressed as analytical functions of the mass and width of the hypothesized resonance.
This approach provides adequate modelling of the invariant mass distribution of the signal for width values up to 10\% of the
resonance mass.

\begin{figure}[!h]
\begin{center}
\subfloat[]{\includegraphics[width=.5\textwidth]{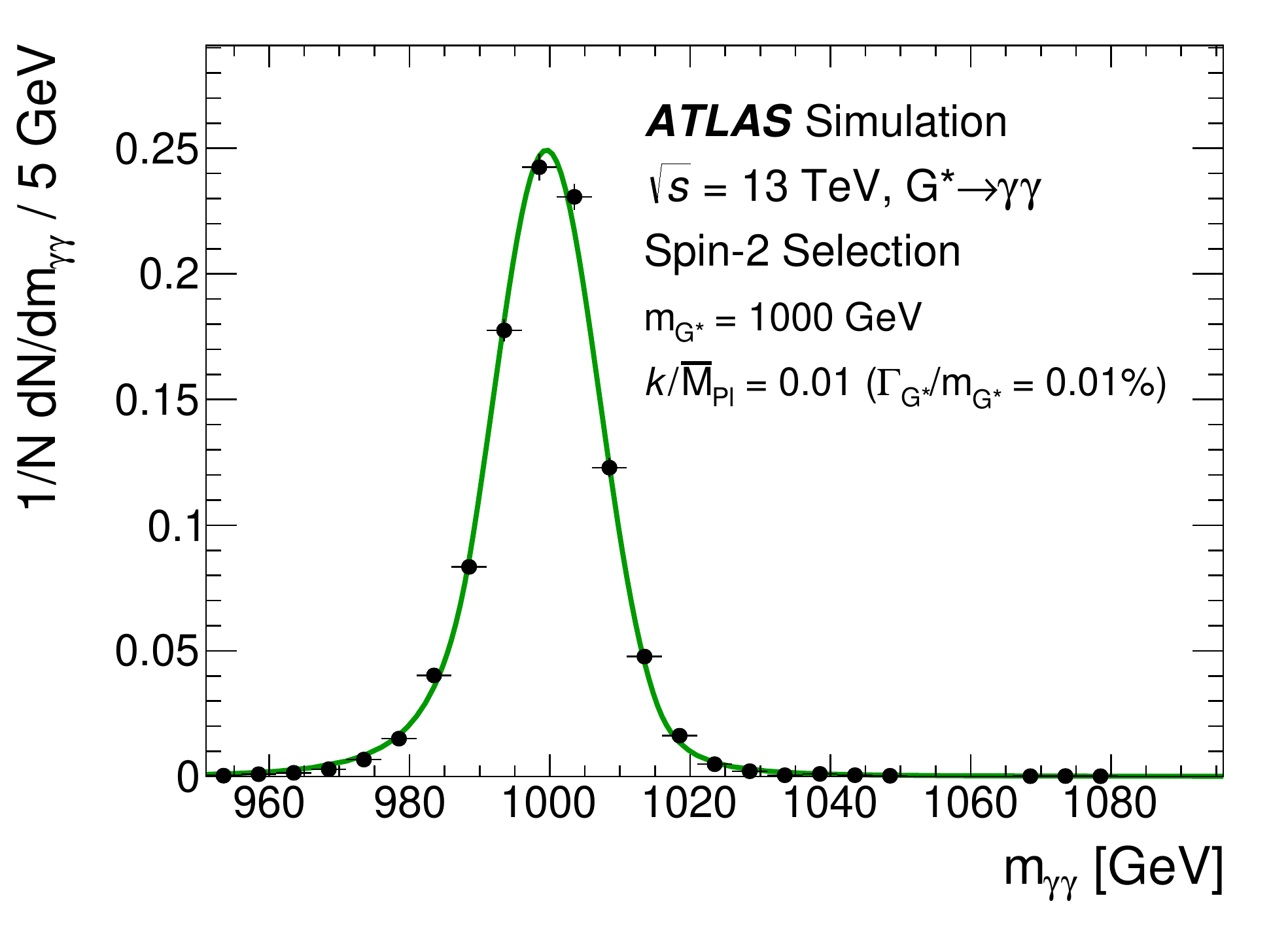}}
\subfloat[]{\includegraphics[width=.5\textwidth]{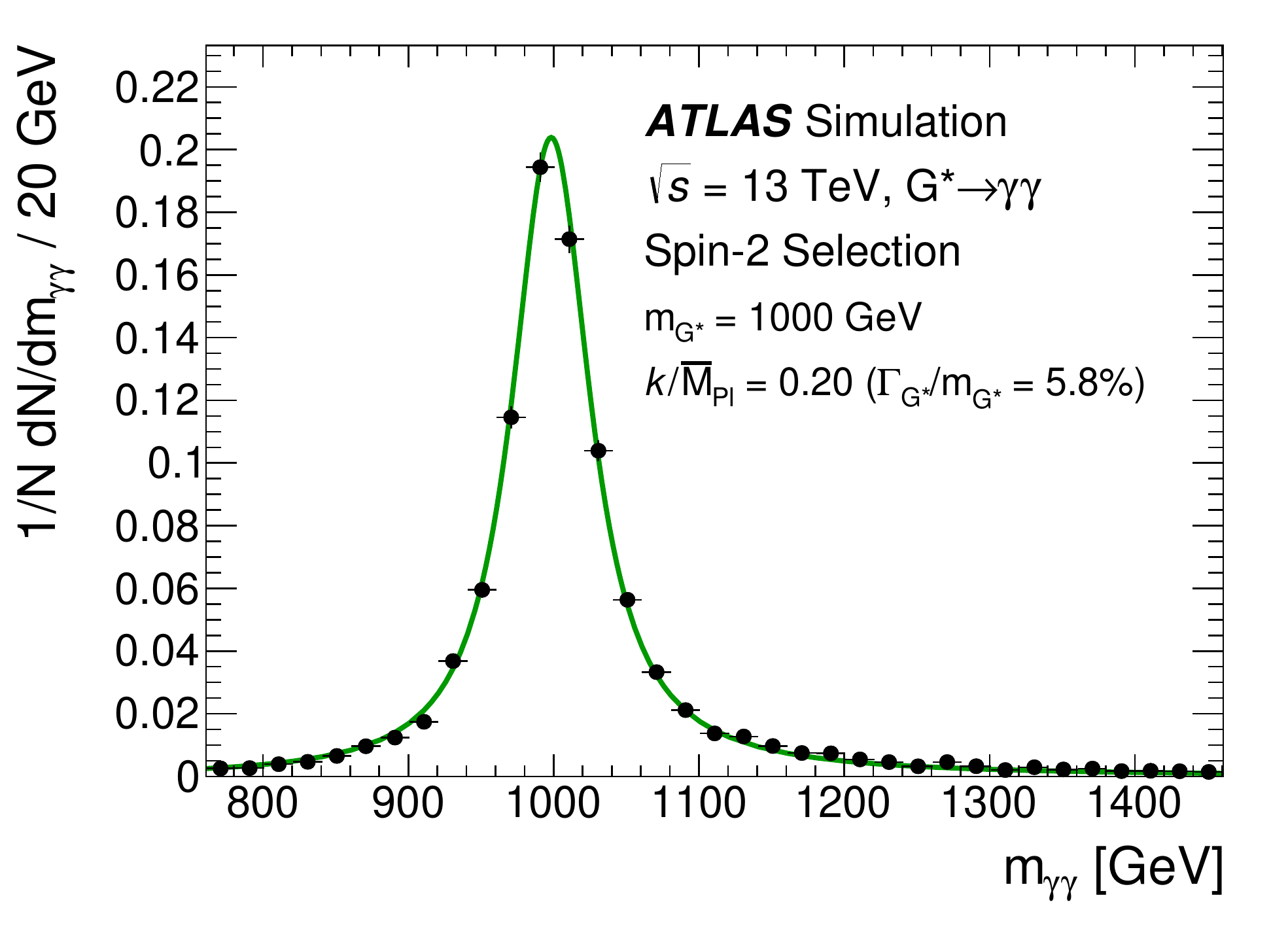}} \\
\subfloat[]{\includegraphics[width=.5\textwidth]{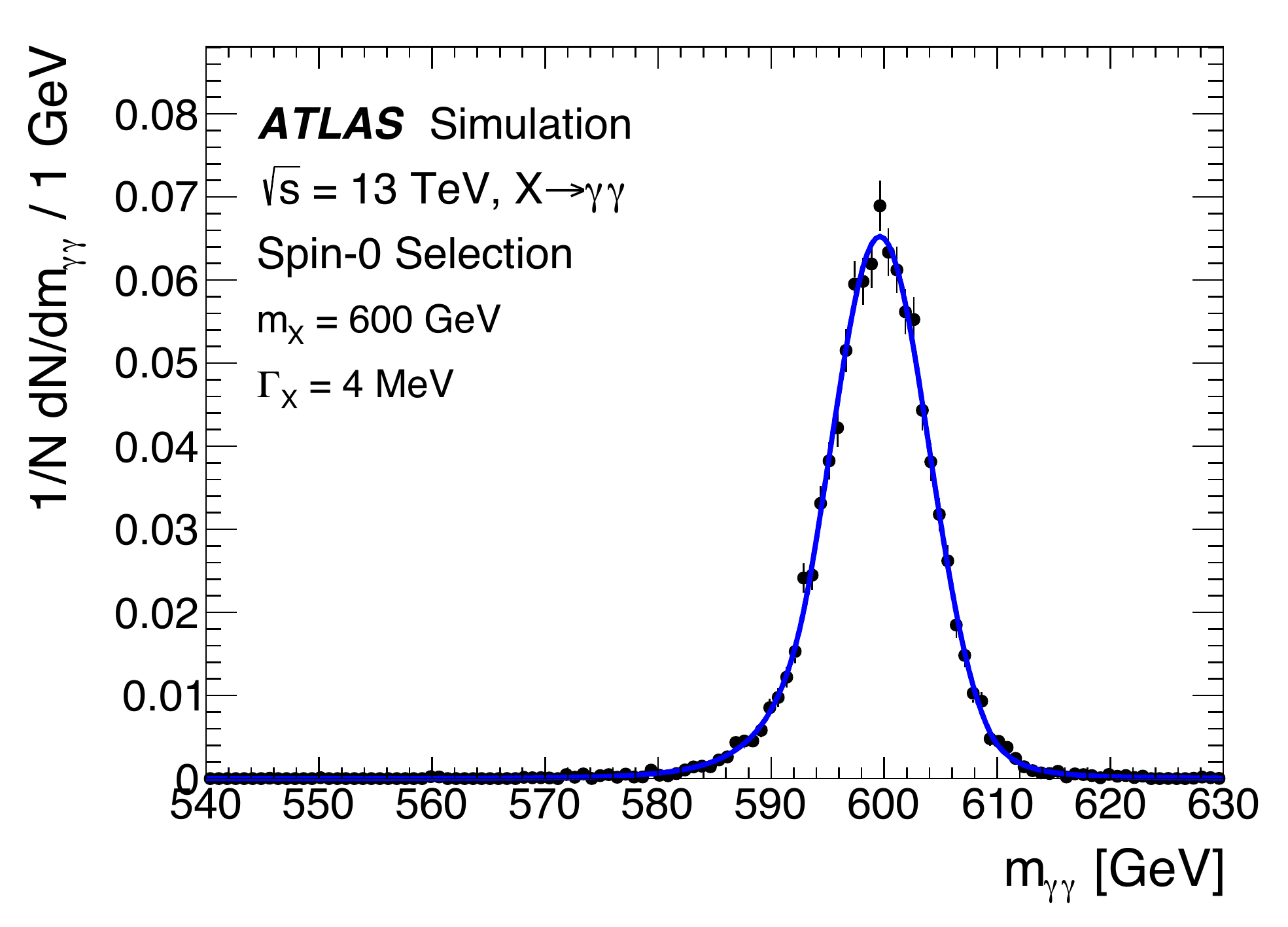}}
\subfloat[]{\includegraphics[width=.5\textwidth]{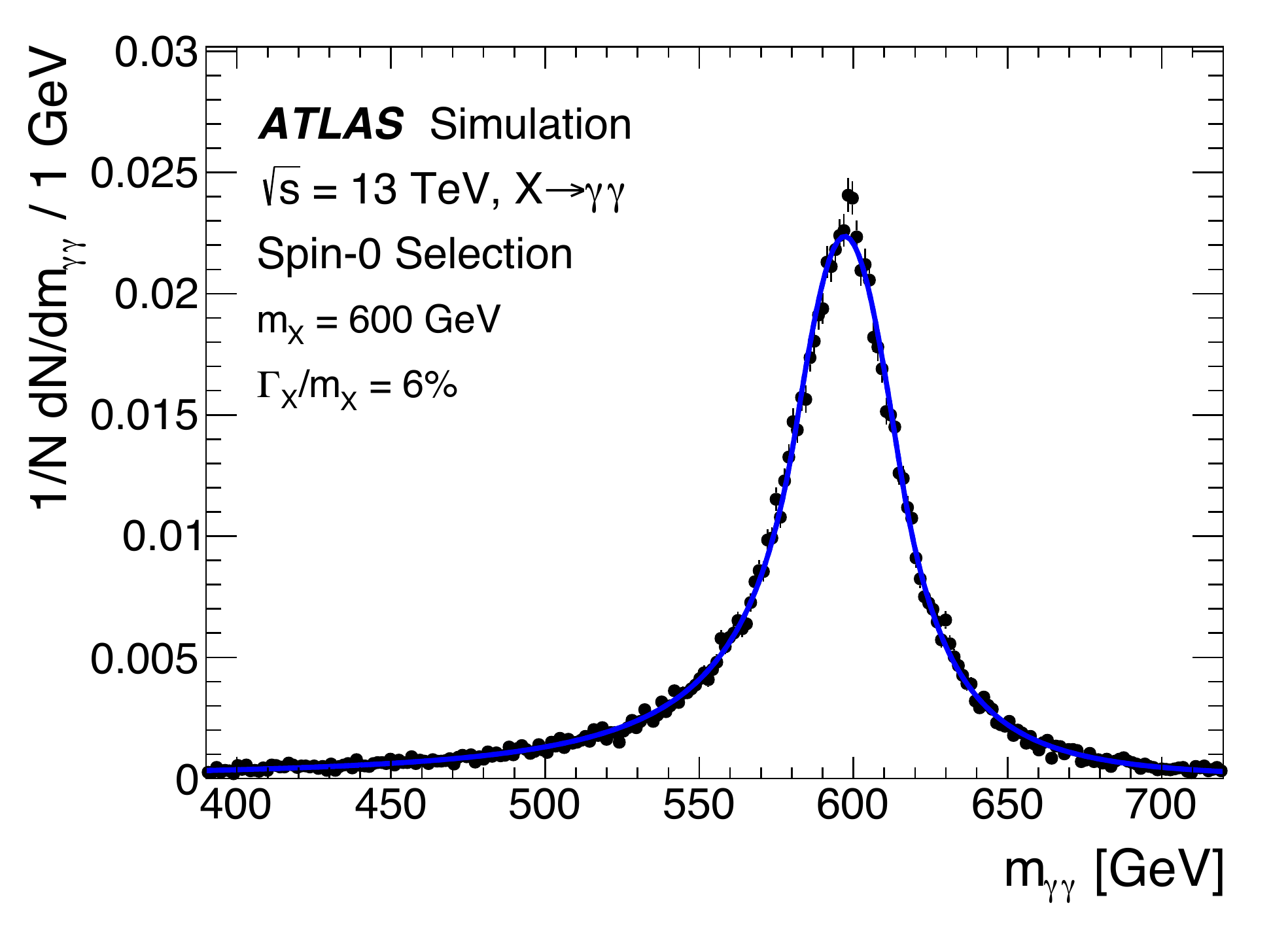}} 
\end{center}
\caption{The $m_{\gamma\gamma}$ distributions for different signal hypotheses for an
RS graviton with a mass of 1000~GeV and (a) \coupling\ = 0.01, (b) \coupling\ = 0.2,
and for a scalar resonance with a mass of 600~GeV and (c) a narrow decay width and (d) with $\Gamma/m$ = 0.06.
A fit is superimposed using the convolution of the graviton mass
line-shape with the detector resolution for the graviton signal case
and using a double-sided Crystal Ball function for the scalar resonance case.}
\label{fig:dscb_fit}
\end{figure}

Figure~\ref{fig:dscb_fit} illustrates the signal modelling 
for a 1000~GeV RS graviton with \coupling\ = 0.01 ($\Gamma_{G^*}/m_{G^*} = 0.01\%$) or 0.2 ($\Gamma_{G^*}/m_{G^*} = 6\%$)
and for a 600~GeV scalar particle
with either a narrow width or a width equal to 6\% of the mass.
For both analyses, a possible bias from the modelling of the signal mass resolution has a negligible impact on the extracted signal yield.


\section{Background estimates}
\label{sec:background}

Two different methods are used to estimate the background contributions to the $m_{\gamma\gamma}$
distribution. 
In the spin-2 search, which aims to reach masses up to 5000~GeV,
the small number of data events at high masses does not effectively constrain the distribution of the invariant mass distribution. 
The shape of the invariant mass distribution of the main diphoton background is thus predicted
using the next-to-leading (NLO) order in QCD \textsc{Diphox}~\cite{Binoth:1999qq} computation, version 1.3.2.
The background from photon+jet and dijet production is added using control samples from the data.
The second approach, more appropriate for the mass range in which there are enough data events
close to  the investigated resonance mass,
is based on using a smooth
functional form, with fully data-driven parameters to model the total background. In this approach,
used for the spin-0 resonance search,
the mass distribution from data is fitted in the range above 150~GeV and the search range for the signal is 200--2000~GeV.

\subsection{Monte Carlo extrapolation approach}
\label{sec:mcbkg}

The background is separated into the diphoton irreducible component and the reducible contributions from photon+jet and dijet events.
To properly normalize each component,
the composition of the data sample in the invariant mass interval from 200~GeV to 500~GeV is determined following a procedure similar to that described
in Section ~\ref{sec:sample_composition}.
The normalized distribution of the total background 
can then be estimated over the full mass range from 200~GeV to 5000~GeV, summing the different background components
with their relative normalizations from the 200--500~GeV range.

The \textsc{Diphox} NLO computation is used to predict the 
shape of the invariant mass distribution of the irreducible diphoton background at the
parton level. 
Kinematic selection requirements corresponding to the analysis selection (\et~$>$~55 GeV, $|\eta|<2.37$) are applied. This computation
includes the contribution of photons produced in the fragmentation of quarks or gluons.
The \textsc{CTEQ6.6M} PDF set~\cite{Nadolsky:2008zw} 
is used and the factorization, renormalization and fragmentation scales
are set to the mass of the diphoton system.
Fully simulated diphoton events  generated with  \textsc{Sherpa}
are reweighted using the ratio of the \textsc{Diphox} and \textsc{Sherpa} 
calculations at the parton level, as a function of the diphoton invariant mass.
The reweighting factor varies by about 20\% over the diphoton mass range from 200~GeV to 5000~GeV.
The uncertainty in the \textsc{Diphox} computation
is estimated by considering the following effects: 
uncertainties in the PDF from variations of the 22 eigenvectors that are provided with
the \textsc{CTEQ6.6M} PDF
(from $\pm$2\% at a mass of 200~GeV, $\pm$35\% at a mass of 3500~GeV
and up to $\pm$140\% at a mass of 5000~GeV on the shape of the normalized  invariant mass distribution),
in the choice of PDF set from a comparison with the \textsc{MSTW2008NLO} PDF set~\cite{Martin:2009iq} (up to $\pm$5\%),
from the photon isolation applied at the parton level in \textsc{Diphox} ($\pm$10\%), 
and from the choice of factorization, renormalization and fragmentation scales used in \textsc{Diphox}
($\pm$5\% in the shape of the normalized invariant mass distribution).

To predict the shape of the photon+jet and dijet backgrounds, control samples where one or two of the photons
fail to meet the tight identification criteria but fulfil looser selections are used. The shape of the invariant
mass distribution in these control samples is fitted with a function of the form
\begin{equation}
f(x) = p_0 \times x^{p_1 + p_2 {\rm log}(x)} \times
\left(1 - \frac{1}{1+\mathrm{e}^{(x-p_3)/p_4}}\right)
\end{equation}
where $x = m_{\gamma\gamma}/\sqrt{s}$ and $p_i$ are free parameters.
The uncertainty in the shape of this background is estimated by varying the identification 
criteria used to select the photons in the control sample. Given the high purity of the selected sample, this is a small
contribution to the total uncertainty.

\begin{figure}[htb]
    \begin{center}
    \includegraphics[width=0.7\textwidth]{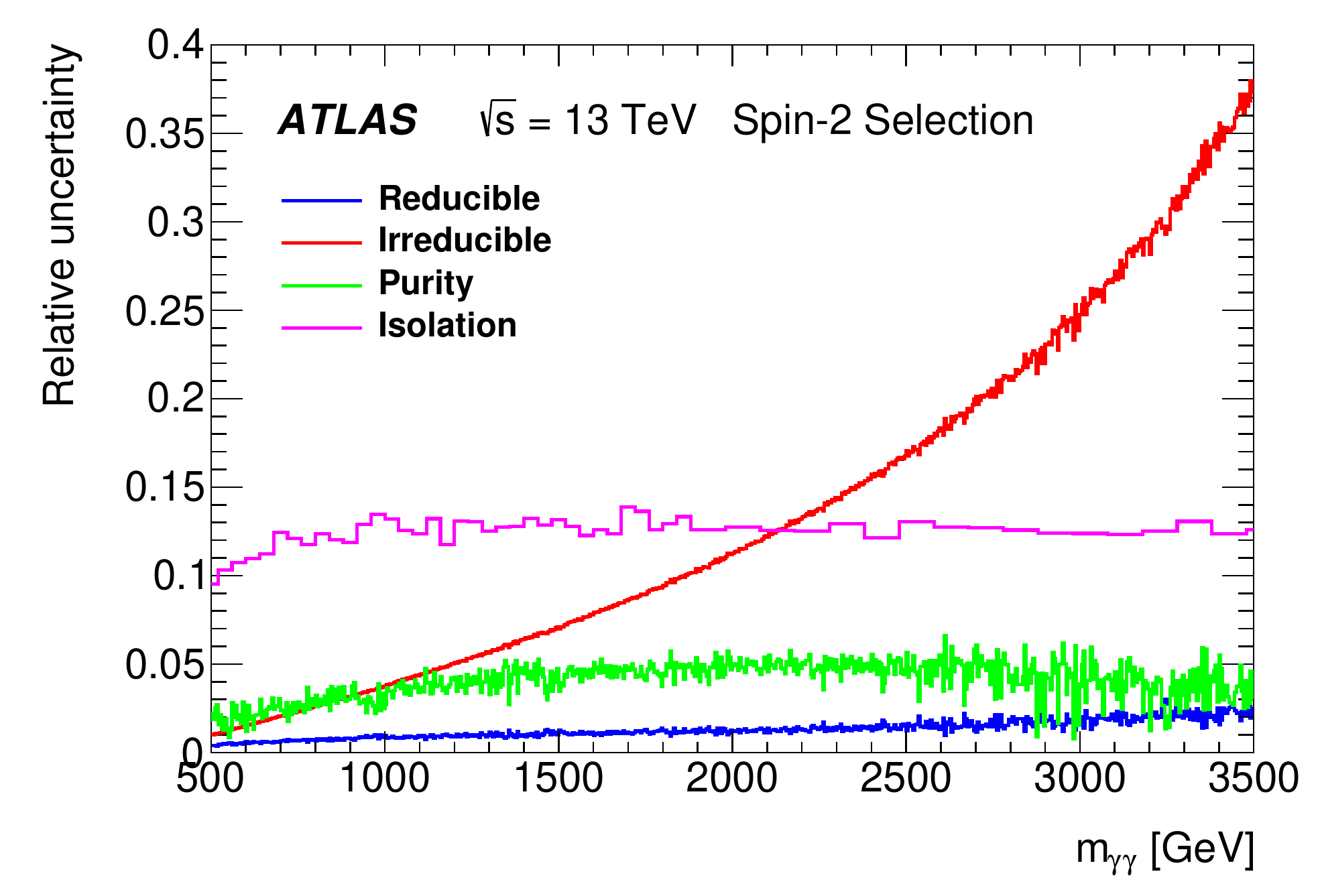}
    \end{center}
    \caption{Relative pre-fit uncertainties in the shape of the $m_{\gamma\gamma}$ distribution of the predicted background for the spin-2 resonance search. 
The uncertainties are shown in the mass range 500~GeV to 3500~GeV.
The reducible background uncertainty corresponds to the uncertainty in the shape of the reducible background component.
The uncertainty in the shape of the irreducible background results from uncertainties
affecting the NLO diphoton computation (parton distribution functions and factorization and renormalization scales).
The uncertainty in the purity corresponds to the impact of the relative normalization of the reducible background compared to the irreducible background.
The uncertainty in isolation results from the
uncertainty due to the choice of parton-level isolation cut in the \textsc{Diphox} NLO computation.}
    \label{fig:spin2_bkgUncertainty}
\end{figure}

The uncertainty in the 
normalized $m_{\gamma\gamma}$ distribution of the total background 
results from uncertainties in both the shape and the relative normalization of each component.
Four independent sources of systematic uncertainty are considered, each of them with an impact varying with the invariant mass
but fully correlated across the full mass range. These sources are the shape of the reducible background,
the relative normalization of the reducible and irreducible backgrounds, the impact of the parton-level isolation requirement
in  \textsc{Diphox} and the effect of the uncertainties in the scales and PDF in the \textsc{Diphox} computation.
Figure~\ref{fig:spin2_bkgUncertainty} shows the evolution
of the uncertainties in the background prediction, before the fit to the invariant mass distribution described in Section~\ref{sec:stat},
as a function of $m_{\gamma\gamma}$ in the region $m_{\gamma\gamma}>500$~GeV,
which is the search range for the spin-2 resonance. 
The sum of these uncertainties is constrained by the fit and reduced significantly compared to the individual components.
At masses larger
than 1000~GeV, the main contribution to the uncertainty  originates from the shape of the irreducible background,
which in turn mostly arises from the PDF uncertainty.
The impact of these uncertainties in the range 200~GeV to 500~GeV
is also taken into account.
In addition, MC statistical uncertainties, which range from $\pm$5\% to $\pm$10\% in a 5~GeV mass interval, are also considered,
uncorrelated between bins.

\subsection{Functional-form approach}
\label{sec:fitbkg}

A family of functions, adapted from those used by searches for new physics signatures in
dijet final states~\cite{Aaltonen:2008dn}, is chosen to describe the shape of the invariant mass distribution:
\begin{equation}
  f_{(k)}(x;b,\{a_{k}\}) = N (1 - x^{1/3})^b x^{\sum_{j=0}^k a_j (\log x)^j} \, ,
  \label{eq:bck_func}
\end{equation}
where $x = m_{\gamma\gamma}/\sqrt{s}$, 
$b$ and $a_{k}$ are free parameters, and $N$ is a normalization factor. The number of free parameters
describing the normalized mass distribution is thus $k+2$.

To validate the choice of this functional form and to derive the corresponding uncertainties, the method
detailed in Ref.~\cite{HIGG-2013-08} is used to check that the functional form is flexible enough to accommodate different
physics-motivated underlying distributions. A large sample of diphoton pseudo-data is
produced using the \textsc{Diphox} NLO computation, where the photon four-vectors are smeared with the
detector resolution, and also with \textsc{Sherpa} generated samples which are then passed through the full detector simulation
and the event reconstruction.  The impact  of the PDF uncertainties on the invariant mass distribution is
also taken into account. The shape of the mass distribution for the reducible photon+jet and dijet backgrounds is estimated
with data control samples selected with one or two of the photons failing the tight identification criteria but fulfilling
a looser set of requirements. These samples are dominated by events with one or two jets misidentified as photons.
As the limited number of data events does not directly allow a precise estimate of the mass distribution 
for masses above 500~GeV, the invariant mass distribution
of these samples is fitted with various smooth functions providing an adequate fit to the data. 
The final pseudo-data set is obtained by summing the diphoton contribution and the smoothed estimate of the photon+jet
and dijet backgrounds.  The bias related to the choice of functional form is estimated as the fitted "spurious" signal~\cite{HIGG-2013-08}
yield in these pseudo-data, which consist only of background events, when performing a signal-plus-background fit for various
signal mass hypotheses. To be selected for the analysis, the functional form is required to have
a fitted "spurious" signal of less than 20\% of the statistical uncertainty in the fitted signal yield
over the full investigated mass range.
Among the forms fulfilling this criteria, the one with the lowest number of degrees of freedom is preferred.
Based on these criteria, the functional form defined in Eqn.~\ref{eq:bck_func} with $k = 0$ is selected. The uncertainty
in the background is estimated from the fitted "spurious" signal.
For a narrow-signal hypothesis, it varies from 7 events at 200~GeV to 0.006 events at 2000~GeV.
For larger hypothesized signal widths, the signal is integrated
over a wider mass range and the background uncertainty is larger, varying from 20 events at 200~GeV 
to 0.04 events at 2000~GeV,
for a hypothesized signal with a relative width $\Gamma/m$ of 6\%.
For a signal mass of 750~GeV, it varies from 0.1 events for a small width to 0.8 events for $\Gamma/m$ of 6\%.

To decide whether a function with increased complexity is needed to describe the data, an $F$-test
is performed.
Two background-only fits, using the simplest validated function and a more complex version using a larger value of $k$,  
are performed on the selected data, binned according to the expected number of background diphoton events. 
A test statistic $F$ is computed from the resulting $\chi^2$ values, and its probability is compared with that expected from a 
Fisher distribution with the corresponding number of degrees of freedom. 
The hypothesis that the additional degree of freedom is not needed to adequately describe the data is rejected if the probability to
have an $F$ value higher than the one observed in the data is less than 5\%.
The tests do not indicate a need for additional degrees of freedom with respect to the simplest function ($k=0$).

\subsection{Comparison of the two methods for the background estimate}
\label{sec:bkgComparison}

The two methods used for the background estimate were compared in the mass range where they are both used.
Because of the different event selection, the functional form used to described the background for the spin-2
search selection, determined following the procedure discussed in Section~\ref{sec:fitbkg}, is different
from the spin-0 search selection and can be expressed as
$f(x;a_0,a_1) = N \mathrm{e}^{(a_0 + a_1 {\rm log}(x)) {\rm log}(x)} $, 
where $x = m_{\gamma\gamma}/\sqrt{s}$,
$a_0$ and $a_1$ are free parameters and $N$ is a normalization factor.
To estimate the background in one selection scenario and for a specific approach, an unconditional signal-plus-background
maximum-likelihood fit with the corresponding signal hypothesis is performed on the data. 
The comparison of the different background estimates is shown in Table~\ref{tab:comparison_bkg} for two different
diphoton invariant mass windows.
For the MC extrapolation method, the statistical uncertainty is directly related to the total number of events in the
data sample and thus is at the level of $\pm$1.5\% for the spin-2 resonance search selection. 
For the functional-form approach, the statistical
uncertainty originating from the uncertainty in the determination of the parameters of the function from the fit to the data is larger,
especially at high masses. 
The systematic uncertainty in the MC extrapolation method originates from the sources of uncertainty discussed in 
Section~\ref{sec:mcbkg}. These uncertainties are then constrained when fitting the background model to the data.
For instance, in the spin-2 search selection for a mass around 750~GeV, the total background uncertainty before
the fit is $\pm$12\%. This is reduced to  $\pm$3.5\% after the fit to the data.
For the functional-form approach,
the systematic uncertainty is given directly by the "spurious" signal uncertainty, which depends on the signal hypothesis being
considered. This systematic uncertainty is lower than that of the MC extrapolation approach. 

At high mass the MC extrapolation gives a smaller uncertainty.  
For masses around 750~GeV they are found to be comparable, while at low mass, 
where the statistical component of the uncertainty using the functional-form approach is small, this approach has a lower total uncertainty.
For all masses, the total uncertainty in the background estimate is significantly smaller than the expected statistical
fluctuations of the background yield in the different signal regions being considered.
Overall there is good agreement between the background predictions from the two methods.

\begin{table}
\begin{center}
\begin{tabular}{lrr}
Investigated signal region   &   \multicolumn{1}{c}{Background from}  & \multicolumn{1}{c}{Background from} \\
                             &  \multicolumn{1}{c}{MC extrapolation}    &   \multicolumn{1}{c}{functional form} \\
\hline
\multicolumn{3}{l}{ $m = 750$~GeV, $\Gamma/m$ = 6\%} \\
720--780~GeV, spin-2 selection &  20.1 $\pm$ 0.3 $\pm$ 0.7 &     21.9 $\pm$ 1.2 $\pm$ 0.4 \\
720--780~GeV, spin-0 selection &   6.7 $\pm$ 0.1 $\pm$ 0.4 &      6.8 $\pm$ 0.7 $\pm$ 0.3 \\
\hline
\multicolumn{3}{l}{ $m = 1500$~GeV, $\Gamma/m$ = 6\%}  \\
1440--1560~GeV, spin-2 selection &  1.14 $\pm$ 0.02 $\pm$ 0.09 &    1.51 $\pm$ 0.27 $\pm$ 0.08  \\
1440--1560~GeV, spin-0 selection &  0.32 $\pm$ 0.01 $\pm$ 0.04 &    0.33 $\pm$ 0.11 $\pm$ 0.04  \\
\hline
\end{tabular}
\end{center}
\caption{Estimated numbers of background events for different signal hypotheses, in a mass window corresponding to $\pm$1.5 times the resolution of the
signal. The results obtained by applying the two methods to compute the background are compared. The statistical and systematic uncertainties in the background estimates
are shown separately with the statistical uncertainty first.}
\label{tab:comparison_bkg}
\end{table}

\section{Statistical procedure}
\label{sec:stat}

The numbers of signal and background events are obtained from maximum-likelihood fits of the $m_{\gamma\gamma}$
distribution of the selected events, 
for ($m_X$, \coupling\ ) hypotheses where a spin-2 resonance from the benchmark RS model is probed,
or for ($m_X$, $\alpha$) hypotheses where the presence
of a spin-0 resonance of mass $m_X$ and width $\Gamma = \alpha m_X$ is probed.

The function used to describe the data can be written as
\begin{equation}
N_{\mathrm{S}} f_{\mathrm{S}} (m_{\gamma\gamma}) + N_{\mathrm{B}} f_{\mathrm{B}} (m_{\gamma\gamma}),
\end{equation}
where $N_{\mathrm{S}}$ is the fitted number of signal events, $f_{\mathrm{S}} (m_{\gamma\gamma})$ 
is the normalized invariant mass distribution for a given
signal hypothesis, $N_{\mathrm{B}}$ is the fitted number of background events 
and $f_{\mathrm{B}} (m_{\gamma\gamma})$ is the normalized invariant mass distribution
of the background events. 
In the spin-2 resonance search, $f_{\mathrm{B}}$ is the sum of the diphoton NLO-based computation and the reducible background contribution.
In the spin-0 resonance search, $f_{\mathrm{B}}$ is described by a functional form with two free parameters. 
The fitted number of signal events is related to the assumed signal cross section times branching ratio to two photons
via the integrated luminosity and the acceptance and detector efficiency correction factors.

Uncertainties in the signal parameterization, the acceptance and detector efficiency correction factors for the signal 
and in the description of the background
shape are included in the fit via nuisance parameters.  
Uncertainties in the signal modelling are constrained with Gaussian or log-normal penalty terms.
In the case of the Monte Carlo approach, discussed in Section~\ref{sec:mcbkg}, the uncertainty
in the shape of each background component and in the relative 
normalization of the different components corresponds to a different nuisance parameter affecting
the total background shape, as illustrated in Figure~\ref{fig:spin2_bkgUncertainty}.
These nuisance parameters are also constrained with Gaussian penalty terms.  In the case of the functional-form approach to describe the background,
the parameters of the function are nuisance parameters without penalty terms,
and the systematic uncertainty in the background description is implemented by the "spurious" signal term, which is constrained
by a Gaussian penalty term and, for a given ($m_X$, $\alpha$) hypothesis, has the same invariant mass distribution as the signal.
This "spurious" signal uncertainty is considered separately for each ($m_X$, $\alpha$) hypothesis without any correlation
between the different investigated mass ranges.

Each fit allows for a single signal component. 
The whole mass spectrum (starting at 150~GeV for the spin-0 resonance search and at 200~GeV for the spin-2 resonance search)
is used for all probed mass hypotheses. 

The local $p$-value ($p_0$) for the compatibility with the background-only hypothesis when testing a given
signal hypothesis ($m_X$, $\alpha$) is based on scanning the $q_0(m_X,\alpha)$ test statistic \cite{Cowan:2010js}:
\begin{equation}
  q_0(m_X,\alpha) = -2 \log \frac{
    L(0,m_X,\alpha, \hat{\hat{\nu}})}{
    L(\hat\sigma,m_X,\alpha, \hat\nu)} ,
\end{equation}
where $\sigma$ is the signal yield and
where the values of the parameters marked with the hat superscript are chosen to unconditionally maximize the likelihood $L$, 
while the value with a double hat is chosen to maximize the likelihood in a background-only fit and $\nu$ 
represents the nuisance parameters which are varied in the fit.
This $p_0$-value is calculated using the asymptotic approximation~\cite{Cowan:2010js}.

Global significance values are computed to account for the trial factors given by the search range.
In a first method, which is used for the results given in this publication,  
a large number of pseudo-experiments are generated
assuming the background-only hypothesis and, for each pseudo-experiment, a maximum-likelihood fit is
performed with the signal mass, width and rate as free parameters, within the search range.
The corresponding $p_0$-value is computed and the global significance is estimated by comparing the minimum $p_0$-value
observed in data to the distribution derived from pseudo-experiments. 
A second method, used as cross-check, is based on the techniques described in Refs.~\cite{Gross:2010qma,Cowan:2010js,Vitells:2011da}.

The expected and observed 95\% confidence level (CL) exclusion limits are computed using 
a modified frequentist approach $CL_s$~\cite{cls} with
the asymptotic approximation \cite{Cowan:2010js}.
Cross-checks with sampling distributions generated using pseudo-experiments are performed for a few signal mass points
and a fair agreement with the asymptotic approximation
is found. The largest differences are of the order of 10-20\% on the cross-section limit for a high mass narrow resonance.

\section{Results}
\label{sec:results}

\subsection{Summary of systematic uncertainties}

\begin{table}
\begin{center}
\begin{tabular}{lcc}
Uncertainty            &    Spin-2 search & Spin-0 search  \\
\hline
Signal mass resolution  & {\multirow{2}{*}{$^{+(30-60)}_{-(20-40)}\%$}}  &  {\multirow{2}{*}{$^{+(40-60)}_{-(30-45)}\%$}} \\
  (mass dependent)  & & \\
\hline
Signal photon identification  & \multicolumn{2}{c}{$\pm(2$--$3)\%$}  \\
 (mass dependent) & & \\
Signal photon isolation    & $\pm(2$--$1)\%$ & $\pm(4$--$1)\%$  \\
 (mass dependent) & & \\
\hline
Signal production process & N/A   &     $\pm(3$--$6)\%$            \\
                          &       &       depending on $\Gamma$  \\
\hline
Trigger efficiency                      & \multicolumn{2}{c}{$\pm0.6$\%}     \\
Luminosity                              & \multicolumn{2}{c}{$\pm5.0$\%}     \\
\hline
\end{tabular}
\end{center}
\caption{Summary of systematic uncertainties in the
signal mass resolution and in the total signal yield (from uncertainties in photon identification, isolation,
process dependence of the reconstruction and identification efficiency $C$ for the spin-0 resonance search, trigger efficiency and integrated luminosity).
For mass-dependent uncertainties the quoted ranges cover the range from 500~GeV (200~GeV) to 5000~GeV (2000~GeV)
for the spin-2 (spin-0) resonance search.}
\label{tab:uncertainties}
\end{table}

The diphoton mass distributions from the two event samples are fitted assuming the background-only or signal-plus-background
hypotheses. Table~\ref{tab:uncertainties} summarizes the input uncertainties related to the modelling of the signal
component in the fit. The uncertainties related to the overall normalization of the signal yield, discussed
in Section~\ref{sec:signalYield}, only affect the limits on the production cross sections, while the uncertainty in
the signal mass resolution affects both the compatibility with the background-only hypothesis and the limits on the production
cross section. For a narrow-width signal, the uncertainty in the mass resolution leads to a $\pm$40\%
relative change in the fitted signal yield at masses near 750~GeV. The impact is much smaller for larger assumed signal decay widths.
For the spin-2 resonance search, the background is estimated using the approach discussed
in Section~\ref{sec:mcbkg}, while the functional-form method described in Section~\ref{sec:fitbkg} is used for the spin-0
resonance search.
The uncertainties in the background estimates, summarized in Table~\ref{tab:comparison_bkg}, affect both the compatibility
with the background-only hypothesis and the cross-section limits. The relative systematic uncertainty in the background
after the fit to the data is typically $\pm$2\% to $\pm$4\%
at masses near 750~GeV depending on the selection and on the method used to compute the background.

\subsection{Compatibility with background-only hypothesis}

\begin{figure}[!h]
\begin{center}
\includegraphics[width=.5\textwidth]{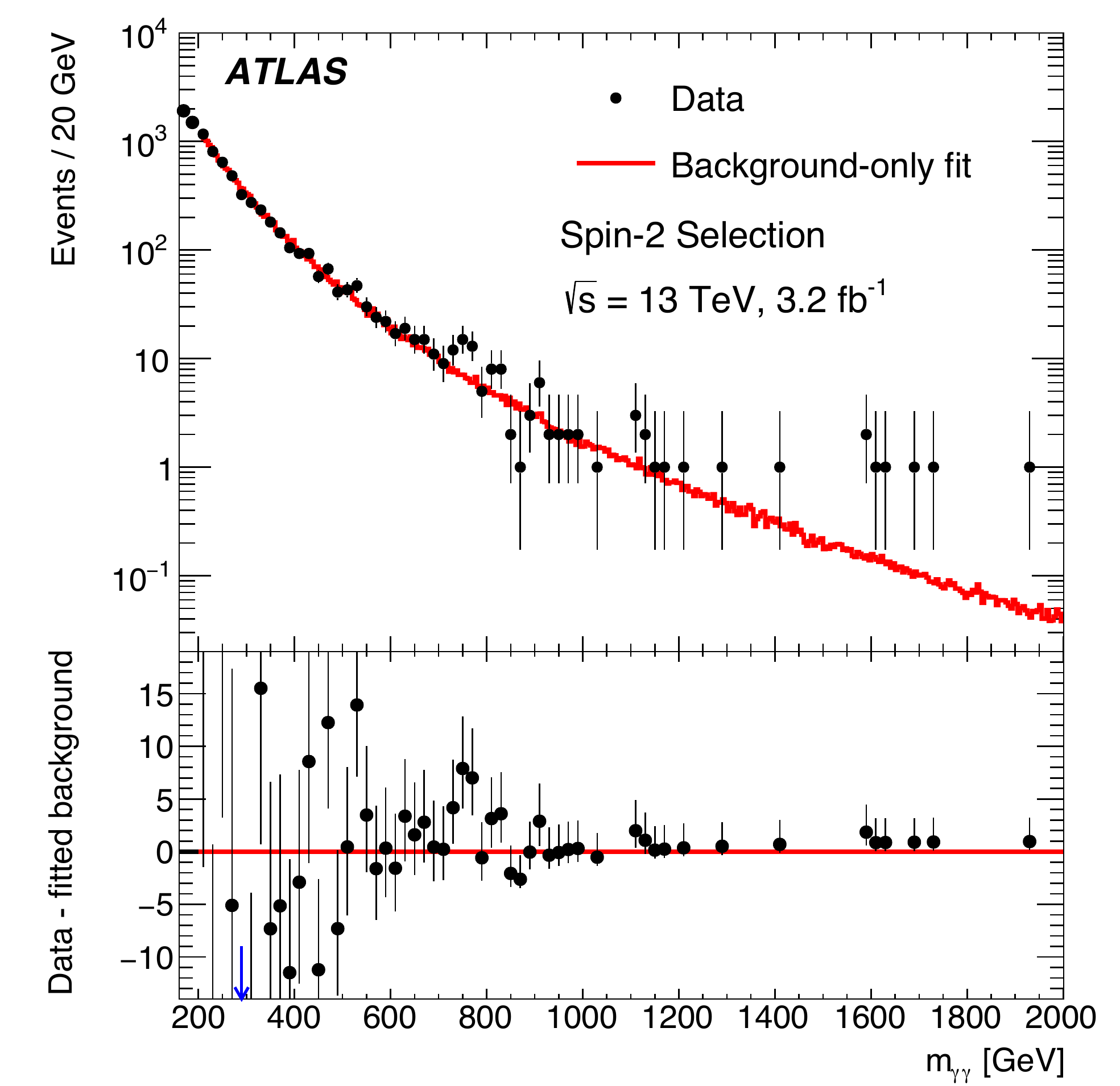}
\end{center}
\caption{Distribution of the diphoton invariant mass for the selection used in the search for a spin-2 resonance, with
the best background-only fit. The difference between the data and this fit is shown in the bottom panel for $m_{\gamma\gamma}$> 200~GeV.
The arrow shown in the lower panel indicates a values outside the range with more than one standard deviation. There is no data event with  $m_{\gamma\gamma}$>~2000~GeV.}
\label{fig:mgg_distributions_spin2}
\end{figure}

\begin{figure}[!h]
    \begin{center}
    \includegraphics[width=.8\textwidth]{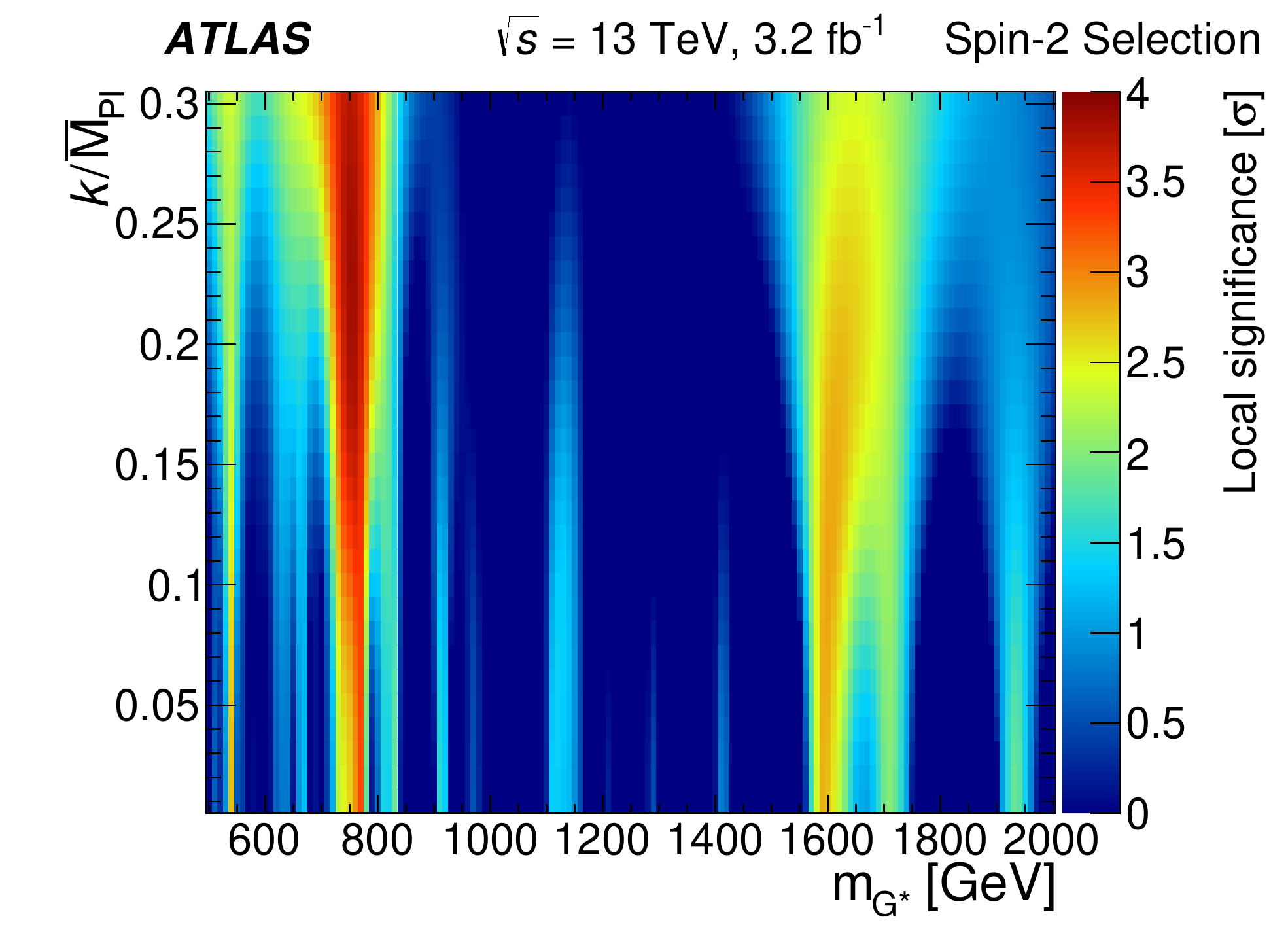}
    \end{center}
   \caption{Compatibility, in terms of local significance $\sigma$,
   with the background-only hypothesis as a function of the assumed signal mass and \coupling\ for the spin-2 resonance search.}
    \label{fig:p0_2D_spin2}
\end{figure}

Figure~\ref{fig:mgg_distributions_spin2} shows the diphoton invariant mass distribution for the events selected in
the spin-2 resonance search together with the best
background-only fit ($N_{\mathrm{S}} = 0$) using the MC extrapolation approach.
Figure~\ref{fig:p0_2D_spin2} illustrates the local compatibility with the background-only hypothesis as
a function of the assumed mass and of the \coupling\ values.

The largest deviation from the background-only hypothesis is observed near a mass of 750~GeV, for
a \coupling\ value of 0.23, corresponding
to a local excess of 3.8 standard deviations.
The width associated with \coupling\ = 0.23 at $m_{G^*}$ = 750~GeV is 57~GeV.
The global  significance  evaluated using the search region of 500--2000~GeV in mass and  0.01--0.3 in \coupling\
is 2.1 standard deviations. The statistical uncertainty from the number of pseudo-experiments is $\pm$0.05 standard deviations.
For \coupling\ = 0.01, corresponding to a narrow width signal, the largest deviation from the background-only
hypothesis corresponds to 3.3 standard deviations local significance at a mass near 770~GeV.
The change in the likelihood ratio between the best signal-plus-background fits with a small \coupling\ value
and \coupling\ = 0.23  corresponds to a difference of 1.3 standard deviations, assuming the asymptotic approximation.

\begin{figure}[!h]
\begin{center}
\includegraphics[width=.5\textwidth]{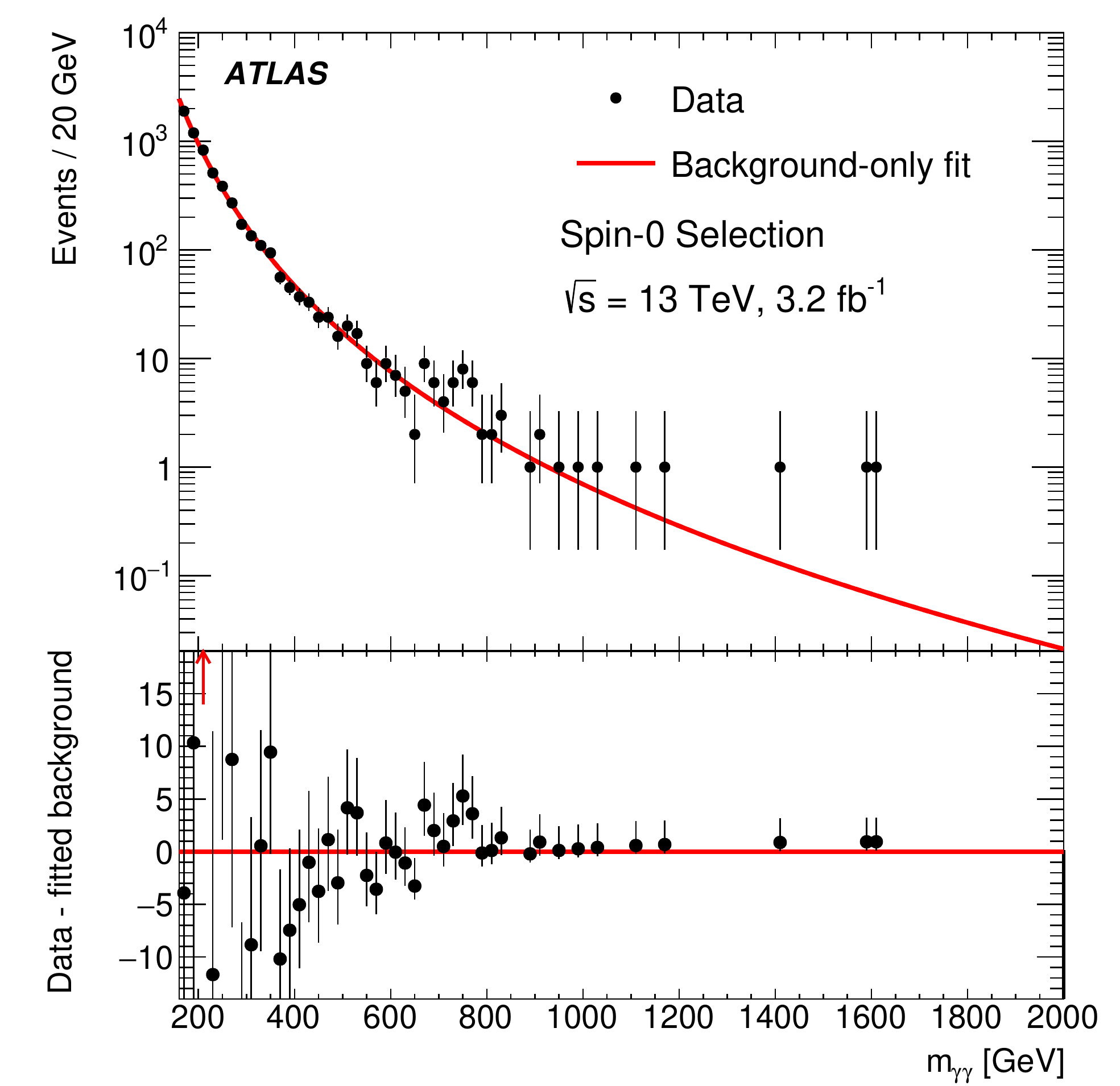}
\end{center}
\caption{Distribution of the invariant mass of the diphoton candidates for the selection used in the search for a spin-0 resonance
with the best background-only fit. The difference between the data and this fit is shown in the bottom panel.
The arrow shown in the lower panel indicates a value outside the range with more than one standard deviation.  There is no data event with  $m_{\gamma\gamma}$>~2000~GeV.}
\label{fig:mgg_distributions_spin0}
\end{figure}

\begin{figure}[!h]
    \begin{center}
    \includegraphics[width=.8\textwidth]{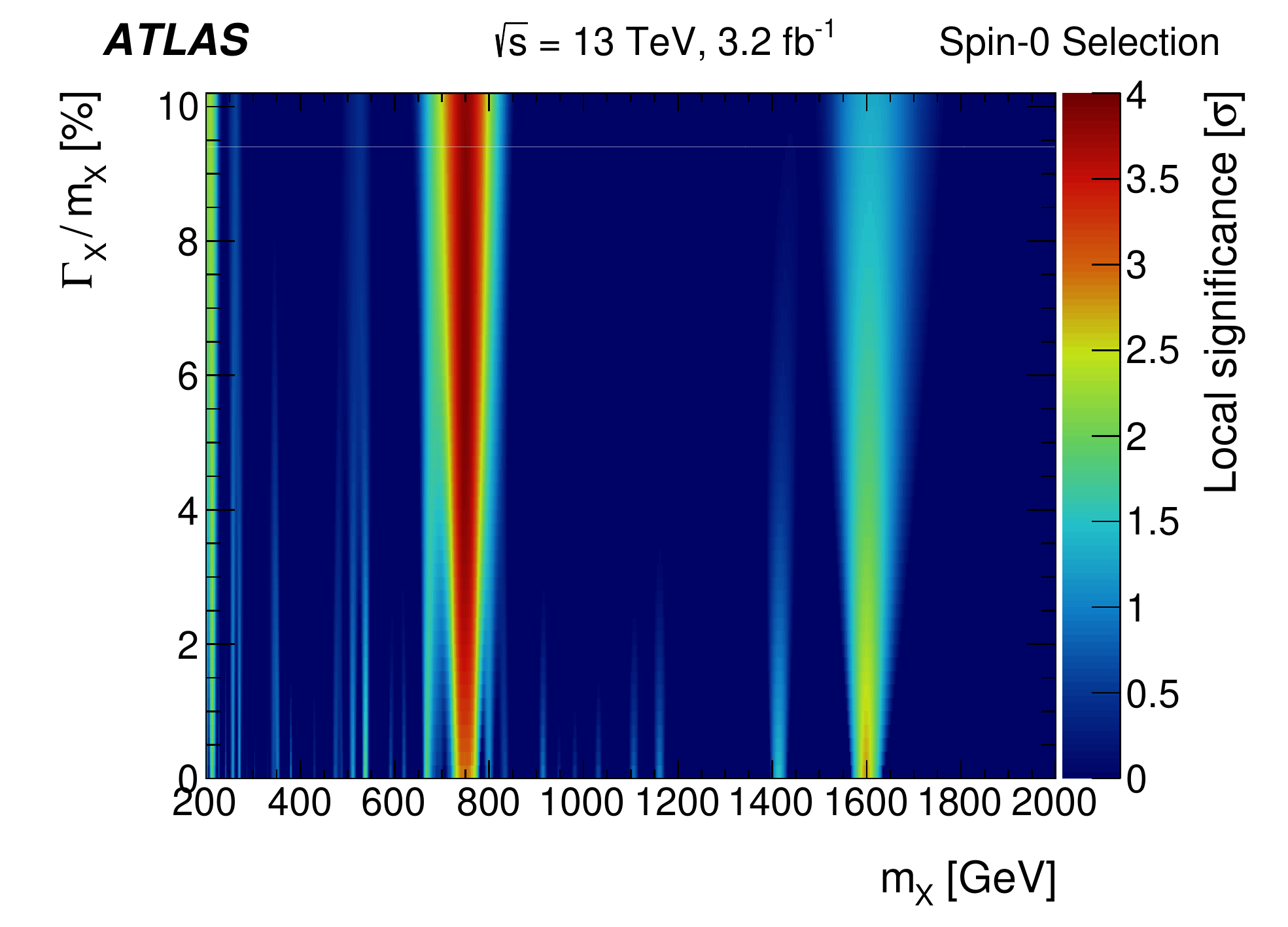}
    \end{center}
   \caption{Compatibility, in terms of local significance $\sigma$,
   with the background-only hypothesis as a function of the assumed signal mass $m_X$ and relative width $\Gamma_X/m_X$ for the analysis optimized for a spin-0 resonance search.}
    \label{fig:p0_2D_spin0}
\end{figure}

Figure~\ref{fig:mgg_distributions_spin0} shows the diphoton invariant mass distribution for the selection optimized
for the spin-0 resonance search together with the best
background-only fit ($N_{\mathrm{S}}$=0) using the functional-form approach.
The compatibility with the background-only hypothesis,
quantified with the local $p_0$-value expressed in standard deviations, is shown in Figure~\ref{fig:p0_2D_spin0} as a function of the
hypothesized resonance mass and width.

As in the spin-2 resonance search, the largest deviation is observed near a mass of 750~GeV. It corresponds
to a local excess over the background-only hypothesis with a significance of 3.9~standard deviations  for a width of 45~GeV.
The impact of systematic uncertainties on the significance of the excess is small, corresponding to a change of
about 0.1~standard deviations in the local significance. 
Only systematic uncertainties related to the background modelling have a non-negligible
contribution to this small difference.
The global significance  evaluated using the search region of 200--2000~GeV in mass and  0\%--10\% in  $\Gamma_X/m_X$
is 2.1 standard deviations. The statistical uncertainty from the number of pseudo-experiments is $\pm$0.05 standard deviations.

If assuming a signal with a narrow width, the largest deviation from the background-only hypothesis is found for a mass near 750~GeV and it
corresponds to a local significance of 2.9~standard deviations. 
The change in likelihood ratio between the best signal-plus-background fits with a narrow width and a width of 45~GeV corresponds to
a 2.5 standard-deviation difference, assuming the asymptotic approximation.

For both selections, the data are well described by the best signal-plus-background fit. Re-binning the mass distribution
to have at least 10 events per bin, a simple $\chi^2$ test-statistic is computed to be 68 for 85 bins 
(65 for 71 bins) for the spin-2 (spin-0) search selection.

The events selected in the spin-0 resonance search constitute a subset of those selected in the spin-2 resonance search,
so the two analyses are not independent.
For a spin-2 signal model, 55\% of the events selected in the spin-2 resonance search also fulfil the selection
criteria of the spin-0 resonance search.
The compatibility of the excesses observed in the two analyses is assessed with a bootstrap statistical procedure, assuming
a common signal. If the spin-2  (spin-0) signal is assumed, the two analyses are compatible within 0.5 (0.2) standard deviations.

\clearpage
\subsection{Kinematic distributions for events with $m_{\gamma\gamma}$ around 750~GeV}

Several cross-checks of the events with invariant masses near 750~GeV were performed and no problem related to the photon
energy measurement or photon identification and reconstruction was found.
A comparison of the properties of the events is made between the events with $m_{\gamma\gamma}$ in the
interval 700--840~GeV and the events in the sideband regions 
with  $m_{\gamma\gamma}$ between 600~GeV and 700~GeV or with 
$m_{\gamma\gamma}$
larger than 840~GeV.
For the selection optimized for the spin-2 (spin-0) resonance search, 70 (31) data events are observed
with  $m_{\gamma\gamma}$ in the interval 700--840~GeV, 77 (29) in the sideband with 600~GeV$< m_{\gamma\gamma} <700$~GeV
and 38 (11) in the sideband $m_{\gamma\gamma}>840$~GeV.
The properties investigated are the number of jets, the transverse momentum
of the diphoton system, the magnitude of the missing transverse momentum (\met) and
the cosine of the angle between the beam axis and the forward-going photon in the Collins--Soper frame~\cite{CollinsSoper} of the 
diphoton system (cos($\theta^*_{\gamma\gamma}$)).
Jets are reconstructed from calorimeter energy clusters using the anti-$k_t$ algorithm~\cite{antikt}
with a radius parameter of 0.4. They are required to have $p_{\rm T} > 25$~GeV and  $|\eta|<4.4$ and,
for jets with $p_{\rm T} < 50$~GeV and  $|\eta|<2.4$,
to fulfil criteria based on tracking information~\cite{ATLAS-CONF-2014-018}
to reject pile-up jets.
The missing transverse momentum is computed as the negative vectorial sum of the transverse momenta
of photons, electrons, muons and jets associated with the diphoton vertex and of a soft term~\cite{ATL-PHYS-PUB-2015-023},
accounting for the
underlying event and soft radiation, which is constructed using tracks from the primary vertex, not associated with other
 identified objects.

\begin{figure}[!h]
    \begin{center}
\subfloat[]{\includegraphics[width=.5\textwidth]{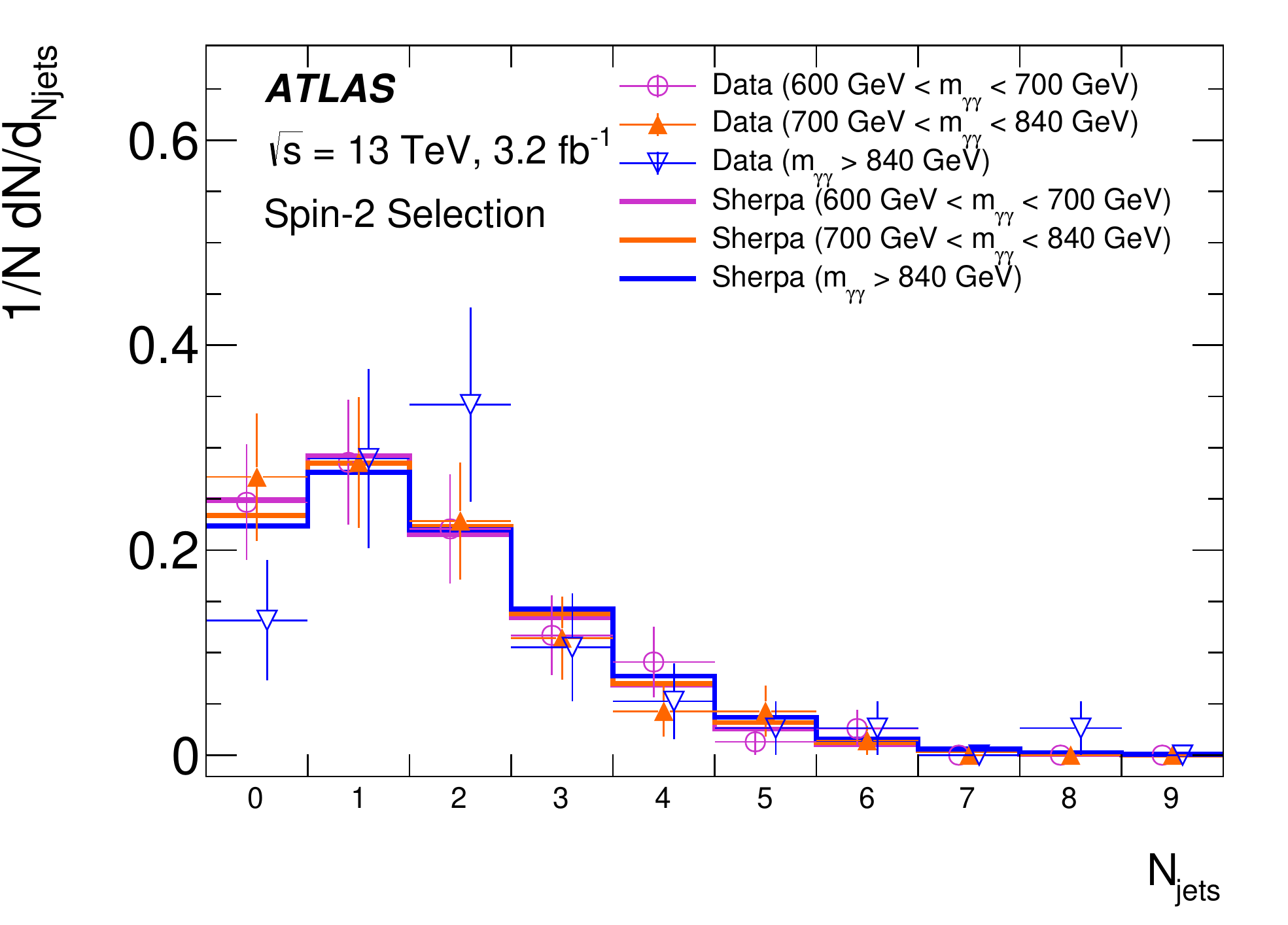}}
\subfloat[]{\includegraphics[width=.5\textwidth]{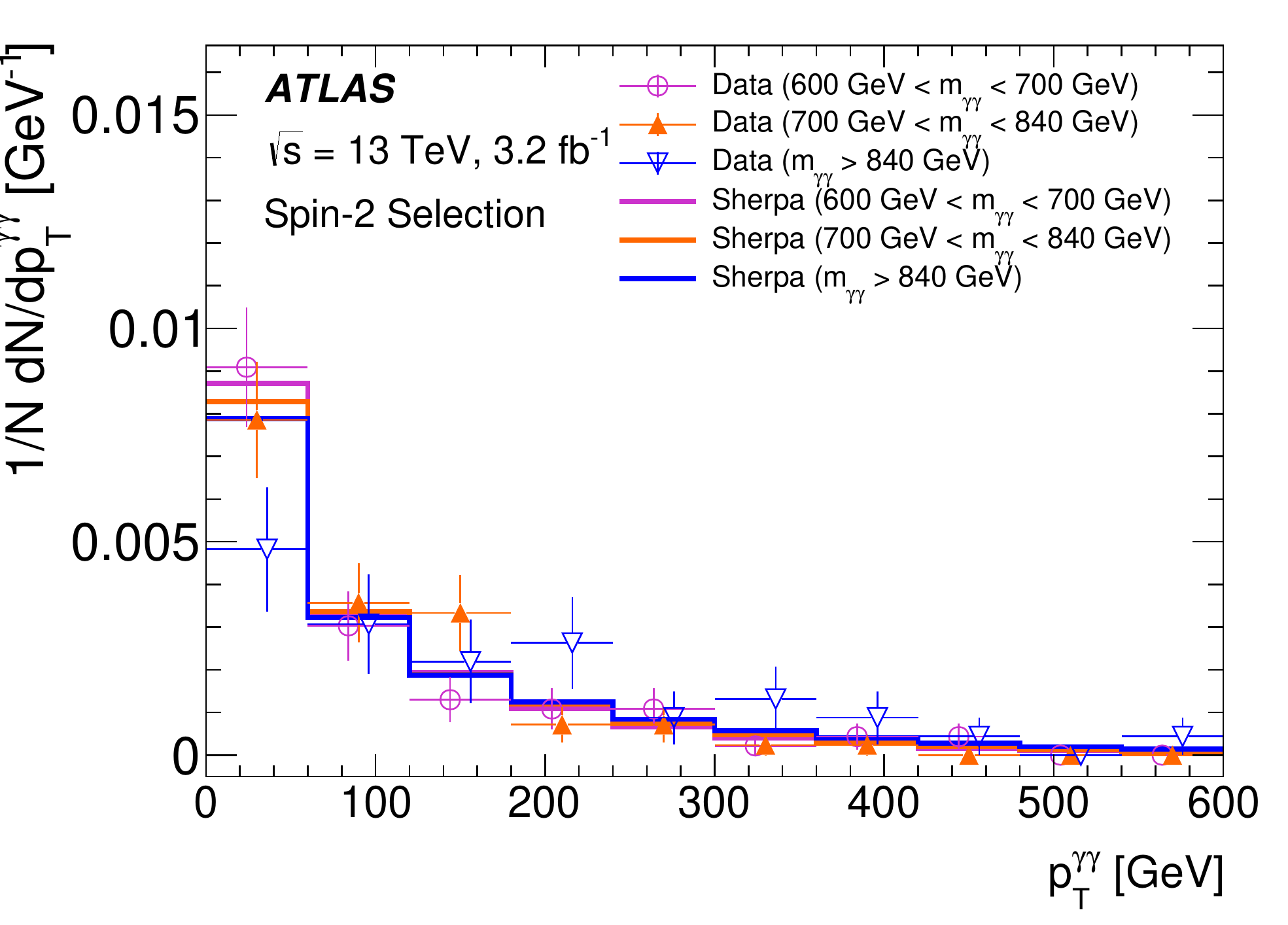}} \\
\subfloat[]{\includegraphics[width=.5\textwidth]{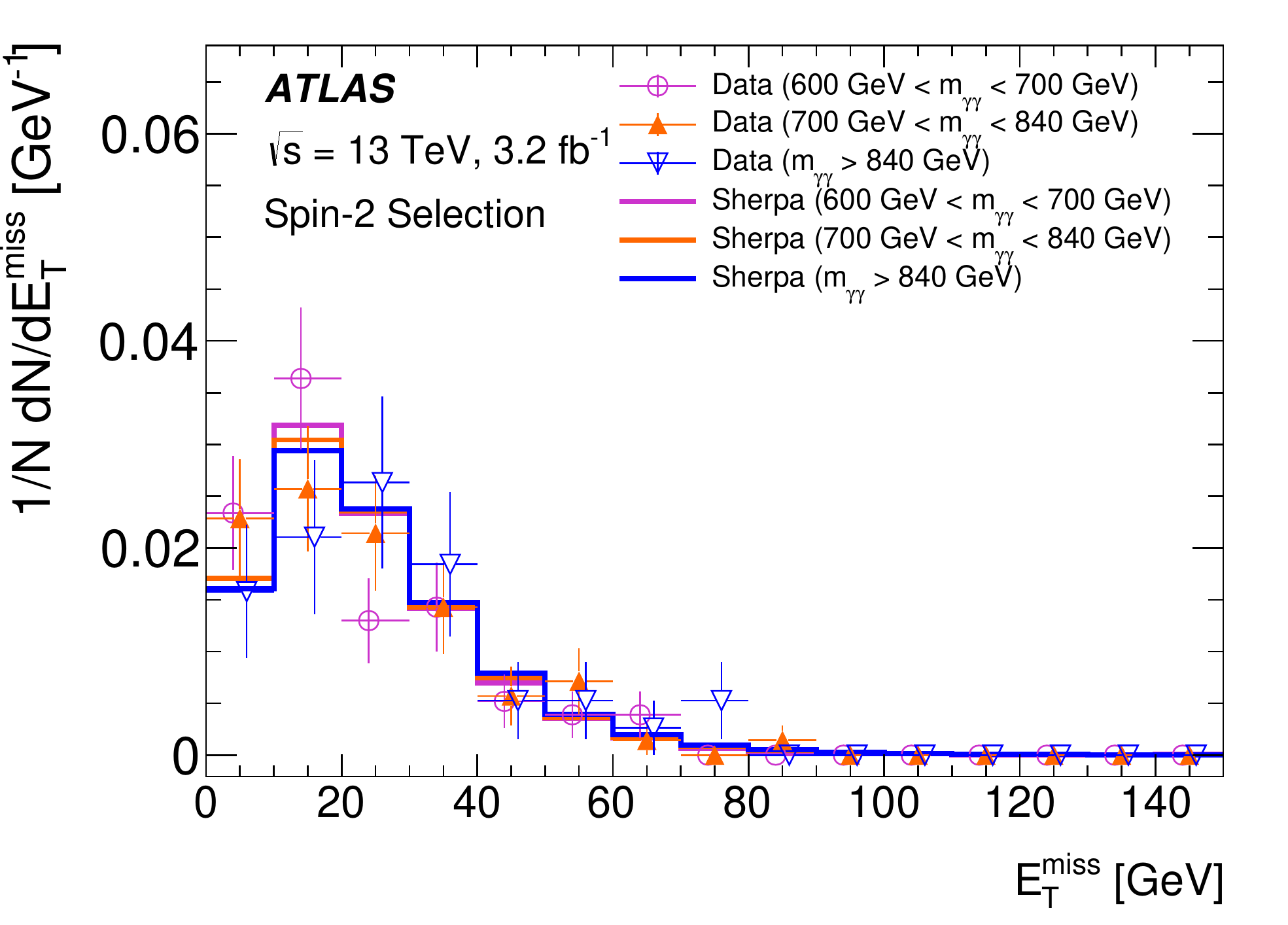}}
\subfloat[]{\includegraphics[width=.5\textwidth]{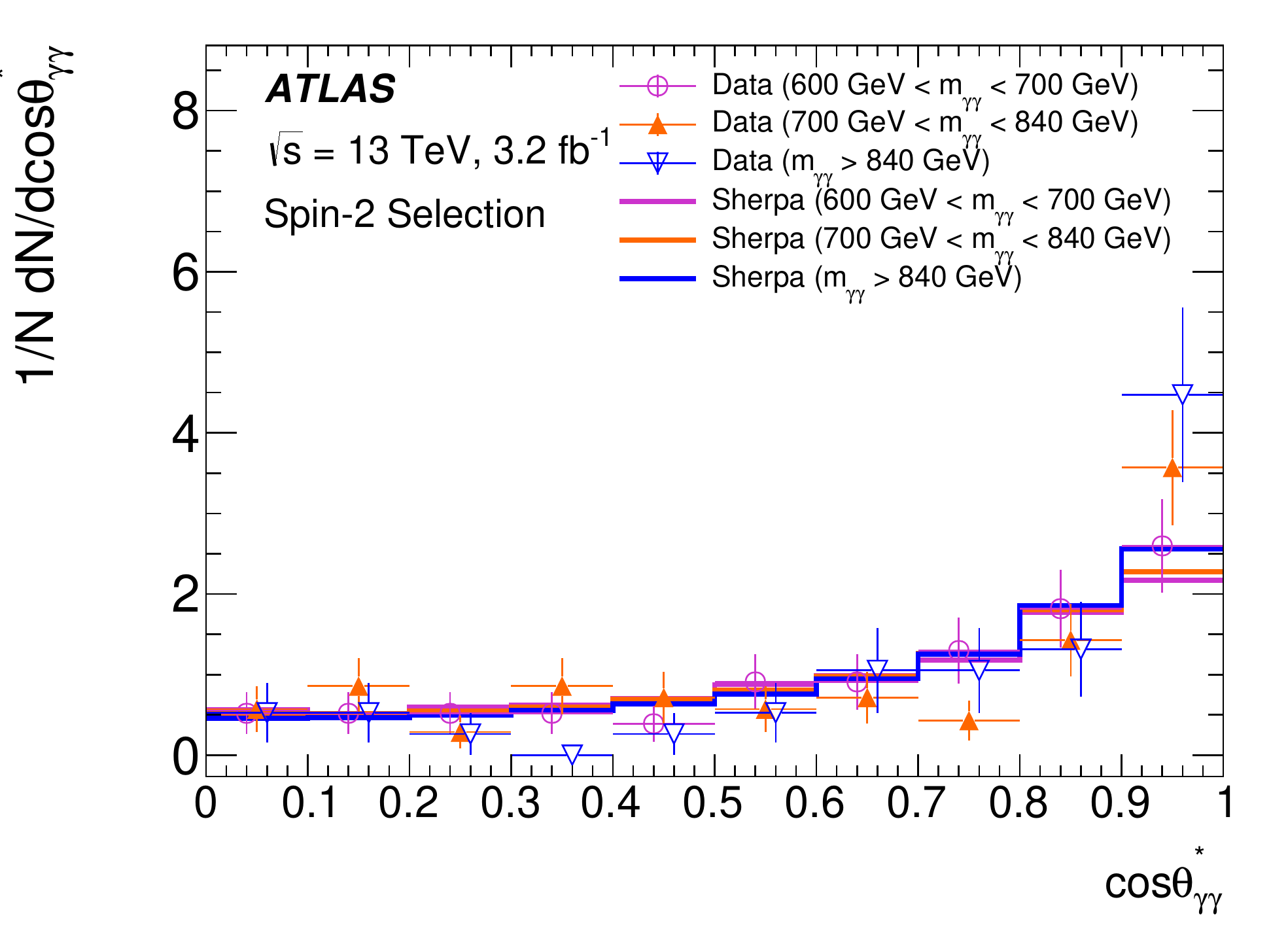}}
    \end{center}
   \caption{Distributions of (a) the number of jets (with $p_{\rm T} > 25$~GeV and  $|\eta|<4.4$) per event, (b) the transverse momentum of the diphoton system,
  (c) the missing transverse momentum and (d) cos($\theta^*_{\gamma\gamma}$) for events
  in the mass interval 700--840~GeV, and the regions 600--700~GeV or $>$~840~GeV, for events fulfilling the analysis optimized for a
  spin-2 resonance search. The  \textsc{Sherpa} predictions for the irreducible $\gamma\gamma$ background are also shown.
 All distributions are normalized to unity.}
    \label{fig:kine_spin2}
\end{figure}

\begin{figure}[!h]
    \begin{center}
\subfloat[]{\includegraphics[width=.5\textwidth]{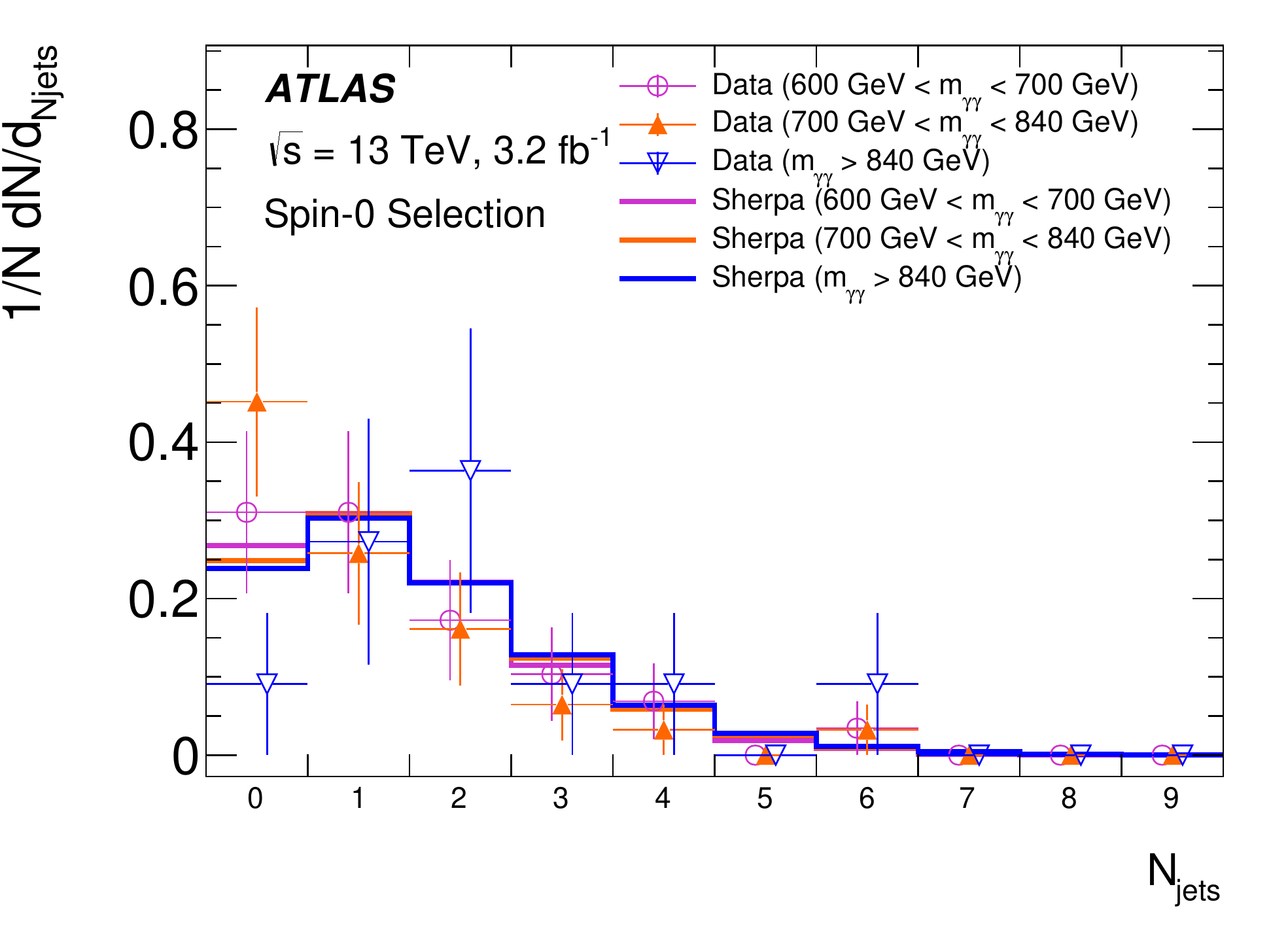}}
\subfloat[]{\includegraphics[width=.5\textwidth]{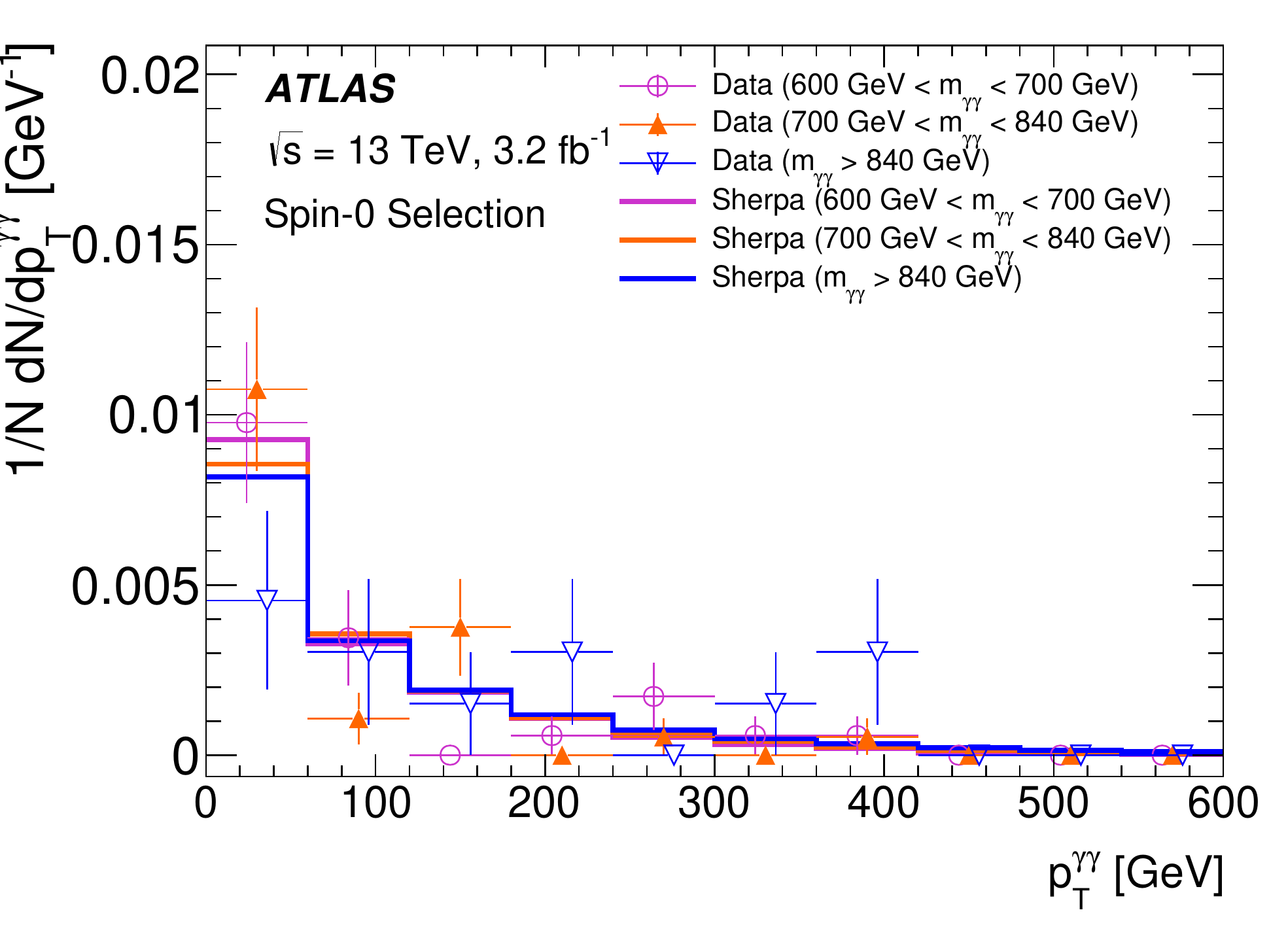}} \\
\subfloat[]{\includegraphics[width=.5\textwidth]{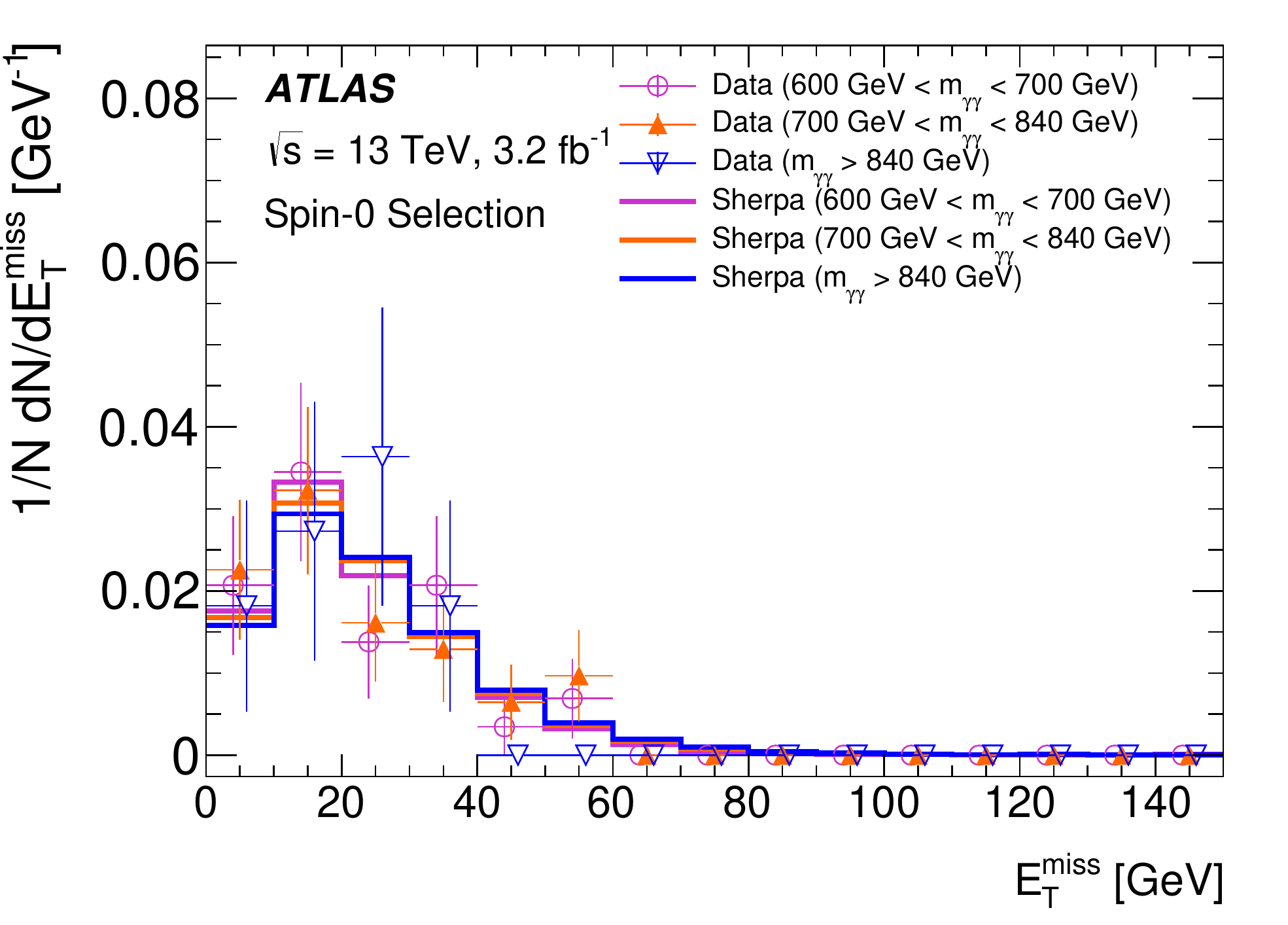}}
\subfloat[]{\includegraphics[width=.5\textwidth]{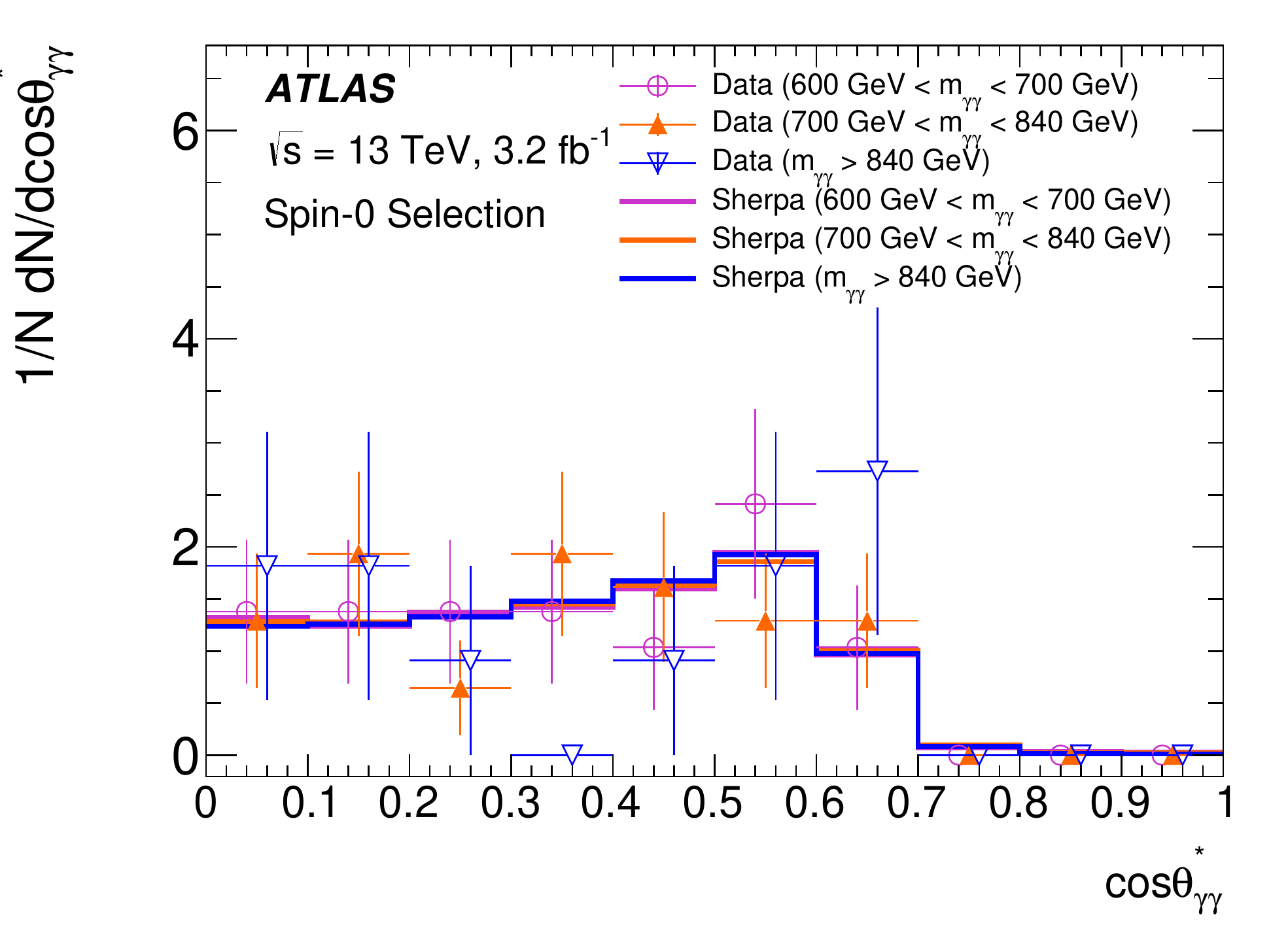}}
    \end{center}
   \caption{Distributions of (a) the number of jets per event  (with $p_{\rm T} > 25$~GeV and  $|\eta|<4.4$) per event, (b) the transverse momentum of the diphoton system,
  (c) the missing transverse momentum and (d) cos($\theta^*_{\gamma\gamma}$)  for events
  in the mass interval 700--840~GeV, and the regions 600--700~GeV or $>$~840~GeV, for events fulfilling the analysis optimized for a
  scalar resonance search. The  \textsc{Sherpa} predictions for the irreducible $\gamma\gamma$ background are also shown.
 All distributions are normalized to unity.}
    \label{fig:kine_spin0}
\end{figure}

Figures~\ref{fig:kine_spin2} and~\ref{fig:kine_spin0} show the distributions of these properties,
for the selections optimized for the spin-2 and spin-0 resonance searches, respectively. The data in the
different mass intervals are also compared to the  \textsc{Sherpa} predictions for the diphoton
background, which represents about 90\% of the total background.
No significant difference between the different mass regions is expected from the
\textsc{Sherpa} predictions of the diphoton background. The data distributions do not show larger differences
than those seen in the simulated sample, with the
distributions in the excess region falling in general between those in the lower and higher mass intervals.
The fraction of events with a jet classified as containing $b$-hadrons, using a multivariate technique~\cite{atlas-btag1,atlas-btag2} with
a $b$-tagging efficiency of 85\%, is about 8\% and is found to be compatible within statistical uncertainties in the different mass regions.
In addition, no electron or muon candidates are found, with $\pt >10$~GeV
and $|\eta|<2.47$ (electrons) or 2.7 (muons) in the events in the three mass regions. The identification selections
of electron and muon candidates
correspond to the medium levels~\cite{ATL-PHYS-PUB-2015-041,Aad:2016jkr}
with loose isolation criteria.

\subsection{Compatibility with 8~TeV data}

The 8~TeV $pp$ collision data recorded in 2012, corresponding to an integrated luminosity of 20.3~fb$^{-1}$, are re-analysed
with a photon energy calibration as described in Ref.~\cite{Aad:2014nim}, which is close to the calibration used for the
13~TeV data. The selections, including the photon isolation and identification requirements, 
are the same as in the original publications~\cite{Aad:2015mna,Aad:2014ioa}
but the spin-0 resonance search is now also performed
at higher invariant masses, up to 2000~GeV, covering the region around 750~GeV.

\begin{figure}[!h]
    \begin{center}
\subfloat[]{\includegraphics[width=.5\textwidth]{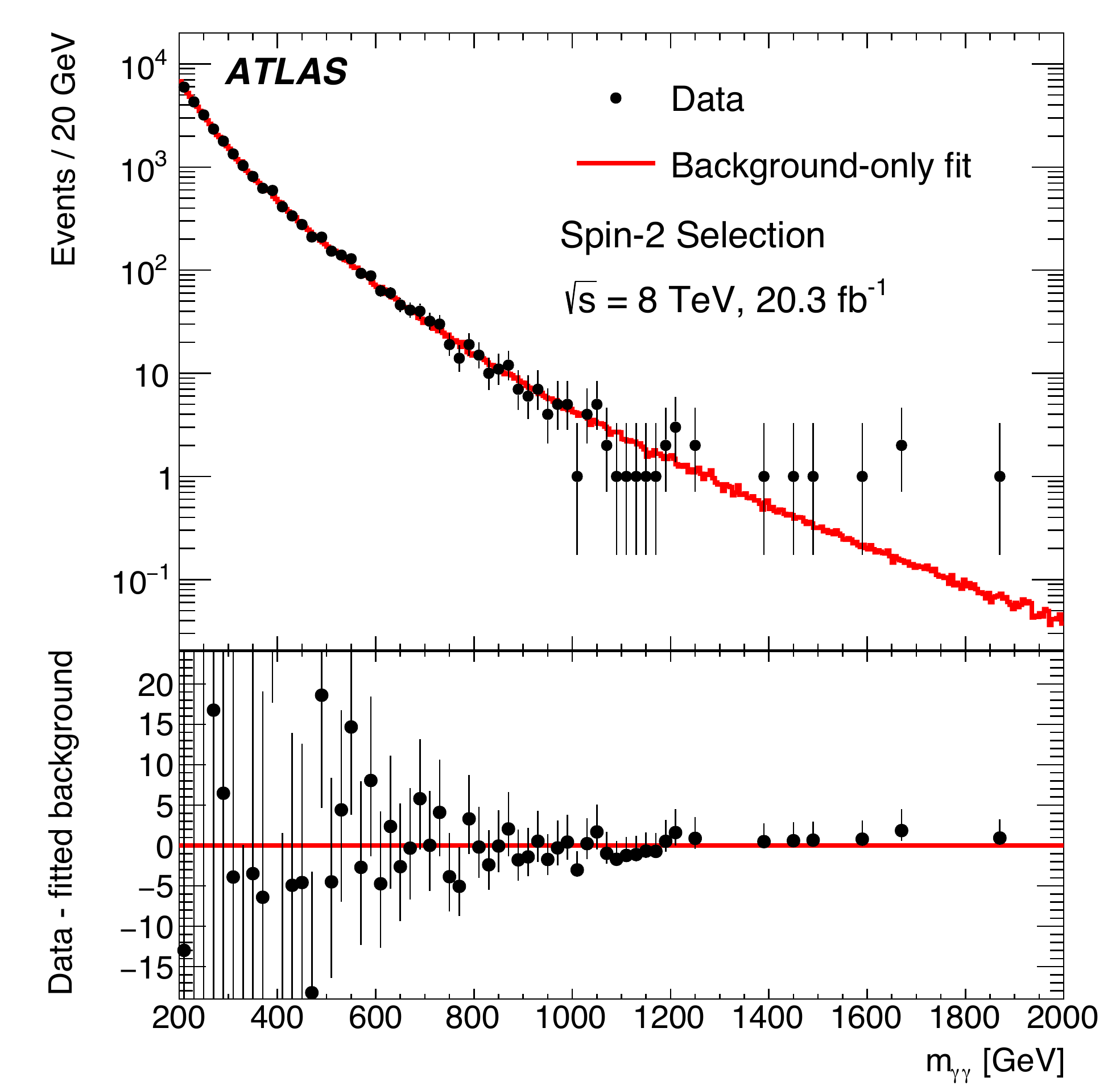}} 
\subfloat[]{\includegraphics[width=.5\textwidth]{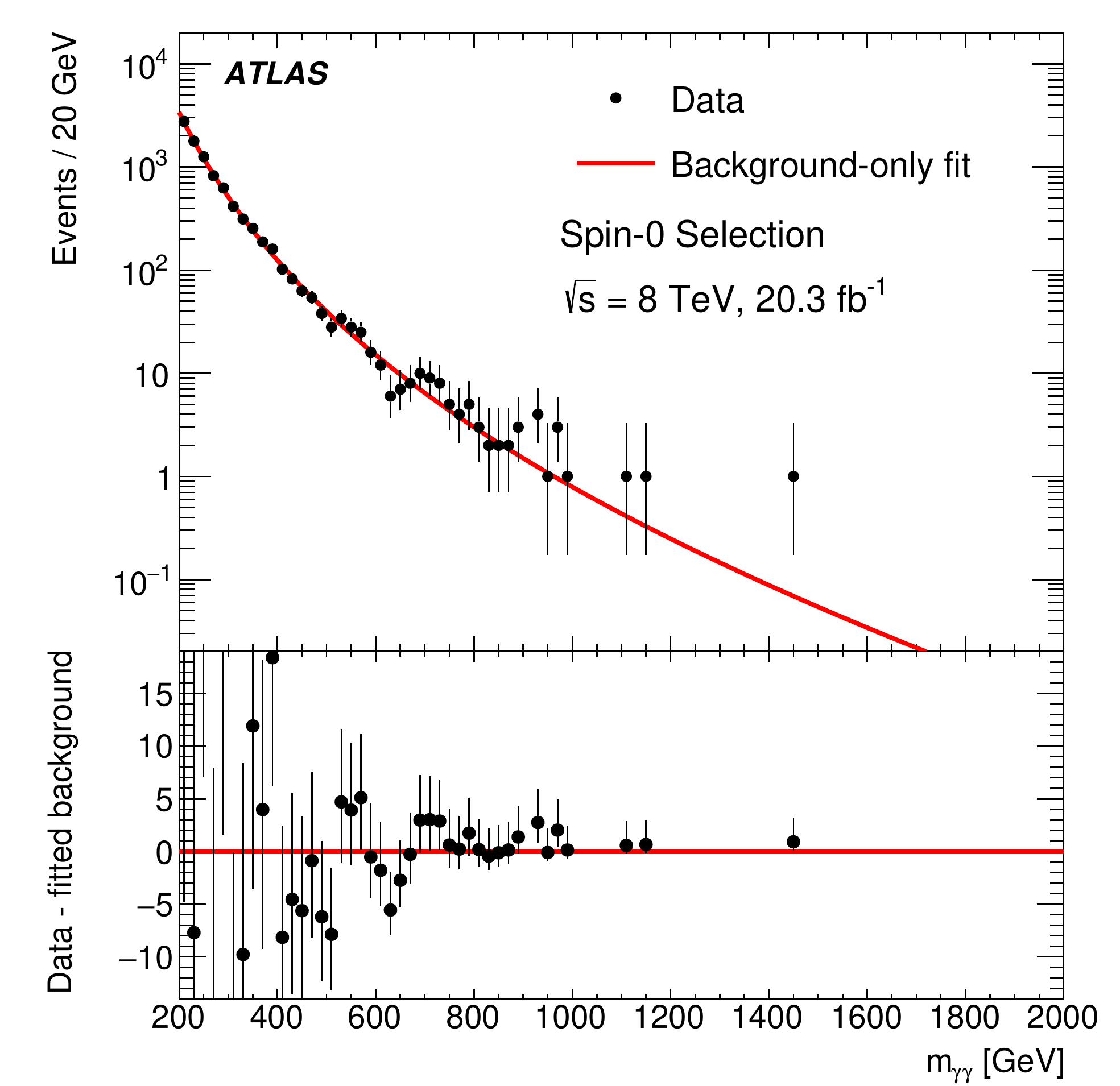}}
  \end{center}
  \caption{Distributions of the invariant mass of the two photons in the 8~TeV data: (a) for the selection optimized
for the search of a spin-2 particle, (b) for the selection optimized for the search of a spin-0 particle. The
data are compared to the best background-only fit.  There is no data event with  $m_{\gamma\gamma}$>~2000~GeV.}
 \label{fig:8TeV}
\end{figure}

The signal and background are modelled following the same methods as described above
and used for the 13~TeV data. The treatment of
systematic uncertainties takes into account the correlations between the two datasets from the common
photon energy calibration procedure.
Figure~\ref{fig:8TeV} shows the invariant mass distributions for the two selections. The number
of selected events with mass above 200 GeV is 24995 (9157) for the spin-2 (spin-0) search selection.

In the search optimized for a spin-2 resonance, no excess is observed in the 8~TeV data above the background prediction for the signal
hypothesis corresponding to the largest deviation from the background-only hypothesis in the 13~TeV data.
This finding is consistent with the published analysis~\cite{Aad:2015mna}.
In the search optimized for a spin-0 resonance, for the  hypothesis of a signal of mass 750~GeV and width  $\Gamma/M =$~0.06,
the 8~TeV data show a small excess corresponding to 1.9 standard deviations.

The consistency of the excess near an invariant mass of 750~GeV between the 8~TeV and 13~TeV datasets is estimated
assuming a common signal model.  For a particle of mass 750~GeV produced as an $s$-channel
resonance, the expected cross section increases by a factor 4.7 for a gluon--gluon initial state and 2.7 for a
light quark--antiquark initial state, as estimated with the MSTW2008NLO or NNLO PDF sets.
The consistency is quantified by adding an extra modifier to the
predicted cross-section ratio, which should be one if the two datasets are consistent, and treating this
modifier as the only parameter of interest in the measurement. The best-fit value of this modifier
corresponds to a smaller cross section at 8~TeV than expected from the 13~TeV excess.

For the analyses optimized for the RS graviton signal model, assuming \coupling\ =~0.2,
the difference between the 8~TeV and 13~TeV results 
corresponds to 2.7 standard deviations  for gluon--gluon production and 3.3 standard deviations
for  quark--antiquark production. For small \coupling\ values, corresponding to a narrow-width RS graviton signal,
the difference is 2.2 (2.4) standard deviations for the gluon--gluon (quark--antiquark) production cross-section scaling.
For the analyses optimized for the spin-0 resonance search, assuming a scalar resonance produced by gluon fusion
with $\Gamma/M =$~0.06, the
difference corresponds to a statistical significance of 1.2 standard deviations
if gluon--gluon production is assumed and 2.1 standard deviations for quark--antiquark production. If 
a narrow-width spin-0 signal is assumed, the difference is 1.5 (2.0) standard deviations for the gluon--gluon (quark--antiquark)
production cross-section scaling.
The tension between the 8~TeV and 13~TeV results in the analysis targeting the RS graviton is similar
if a spin-0 signal model, instead of an RS graviton signal model, is used to compare the results.

\subsection{Cross-section limits}

For both analyses, limits on the cross section times branching ratios are derived. For the spin-0 case,
they are interpreted in a nearly model-independent way in terms of the fiducial cross section, 
defined as the product of the cross section times the branching ratio to two photons 
within the fiducial acceptance defined in Section~\ref{sec:signalYield}. 
Since for the spin-2 case a larger model dependence exists and the analysis is performed for a
specific benchmark model of a spin-2 graviton, 
limits on the total cross section times branching ratio to two photons are given assuming specific model parameters.

\begin{figure}[!h]
\begin{center}
\subfloat[]{\includegraphics[width=.5\textwidth]{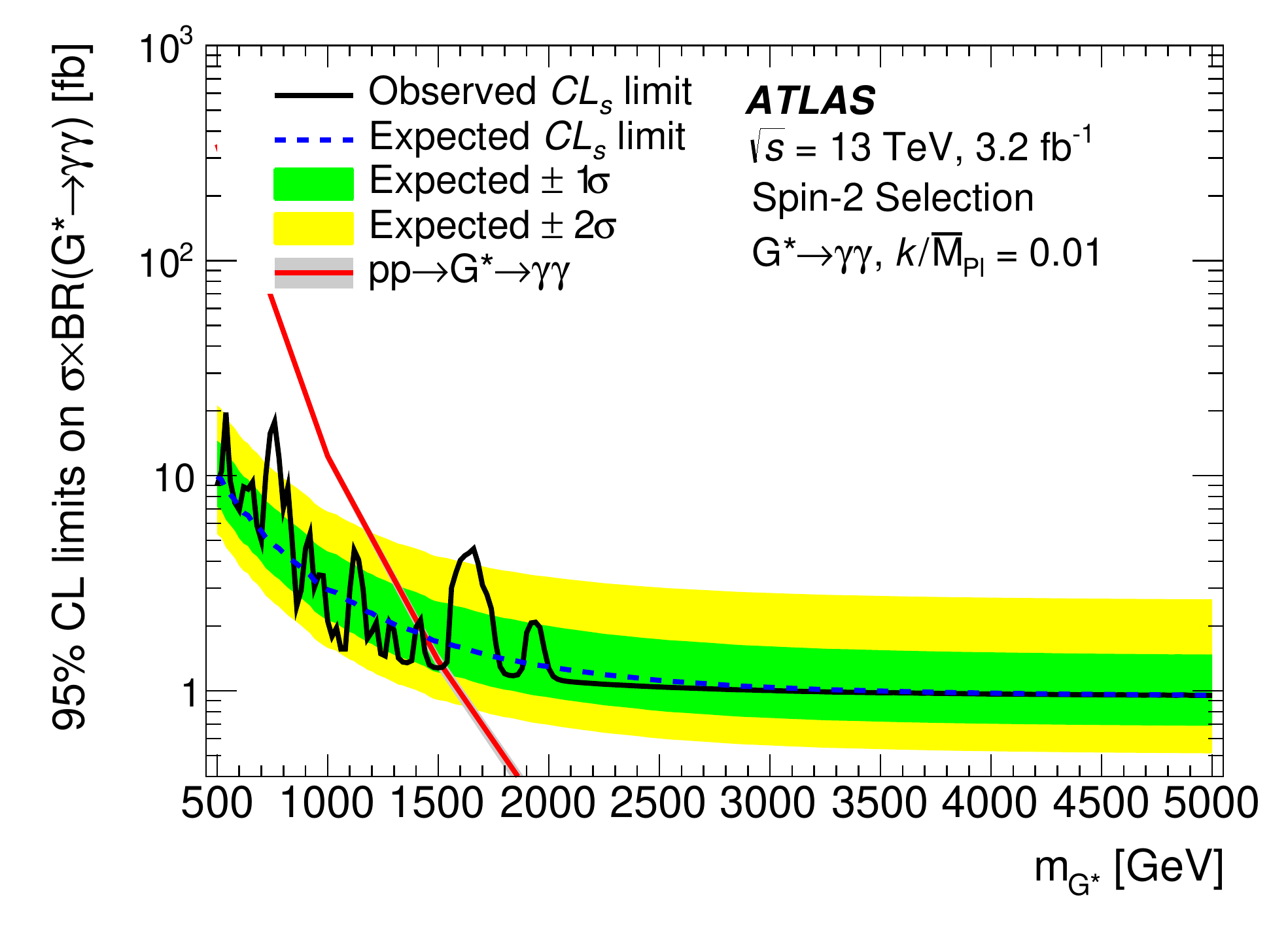}}
\subfloat[]{\includegraphics[width=.5\textwidth]{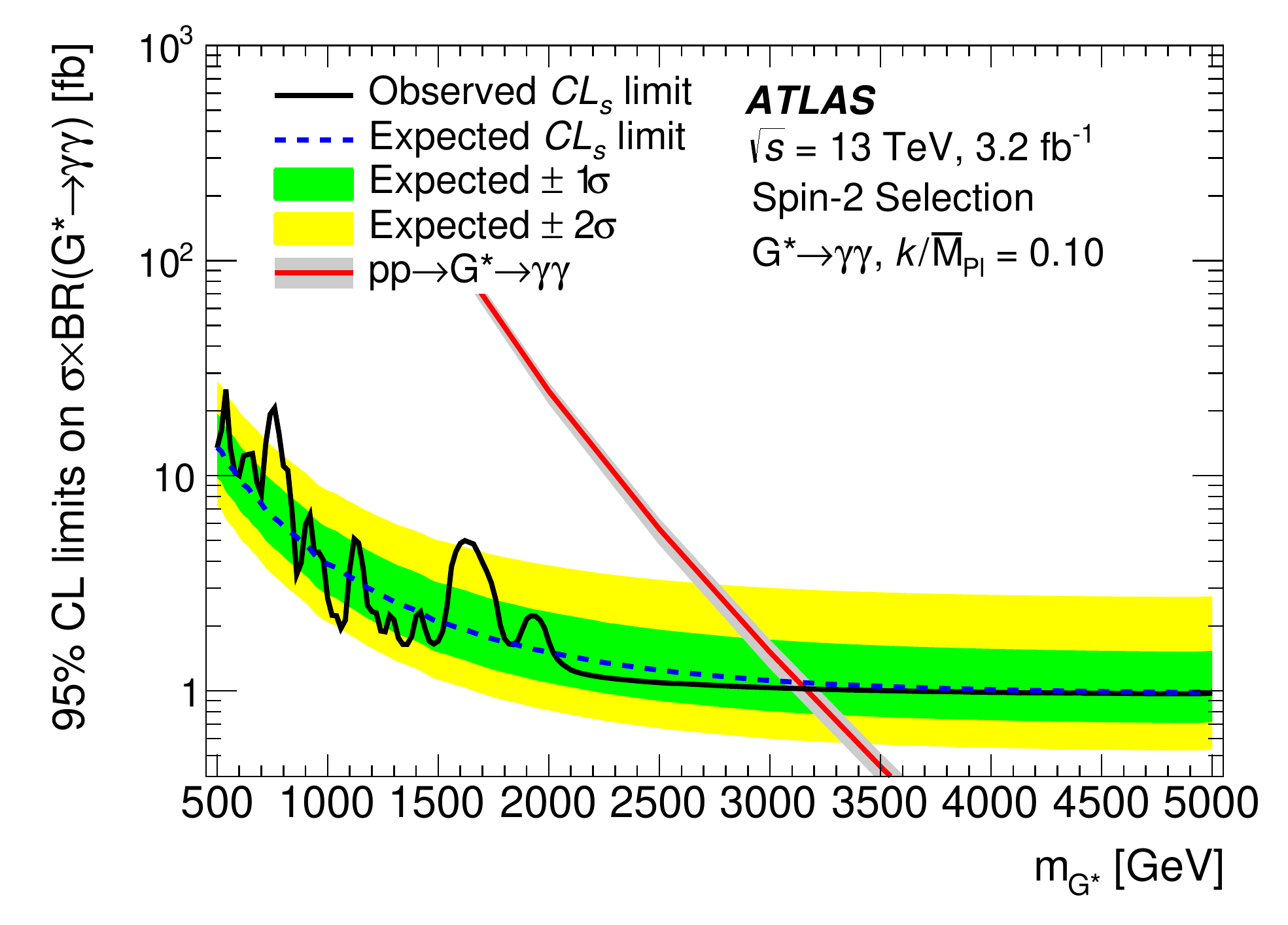}} \\
\subfloat[]{\includegraphics[width=.5\textwidth]{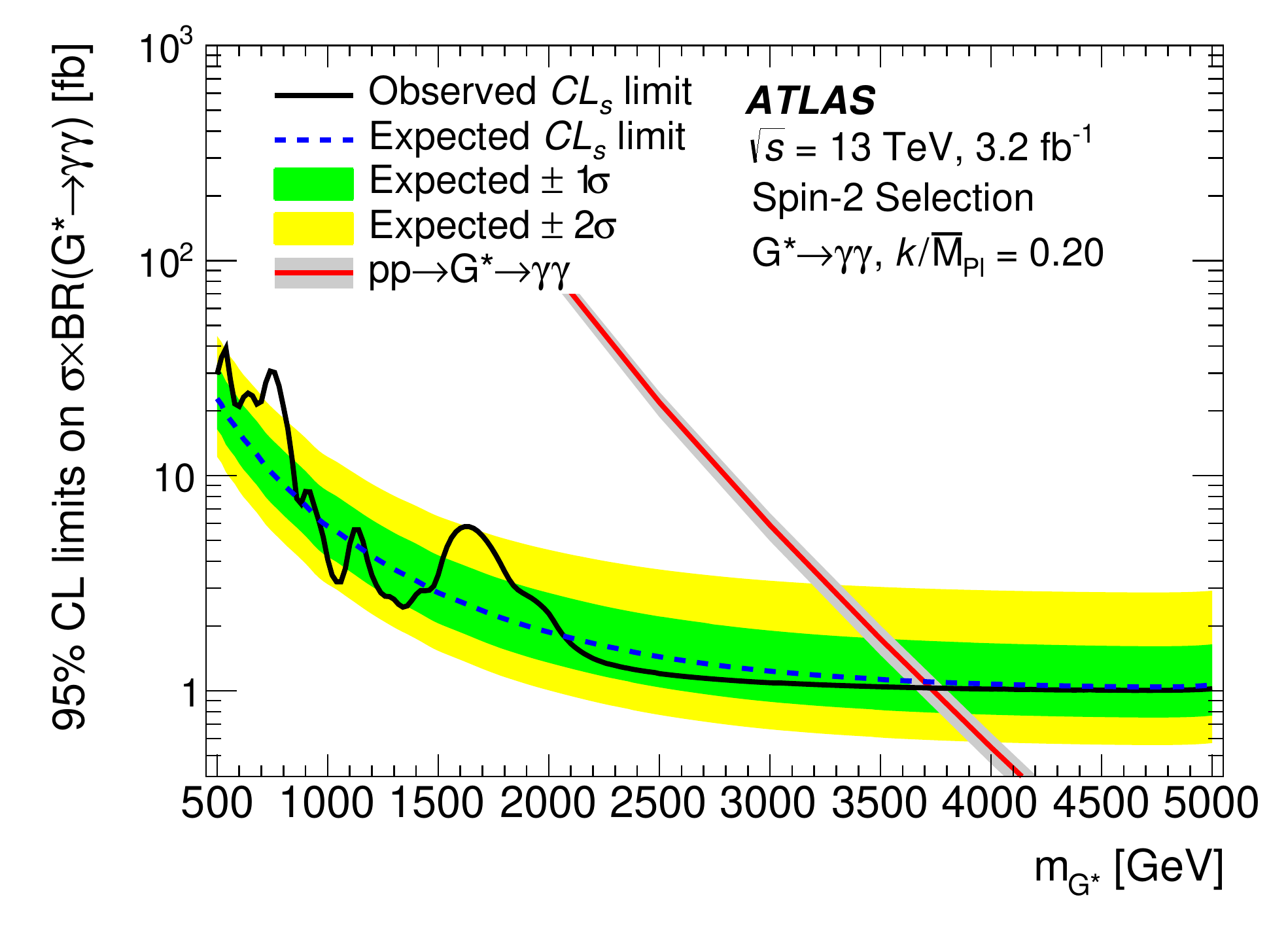}}
\subfloat[]{\includegraphics[width=.5\textwidth]{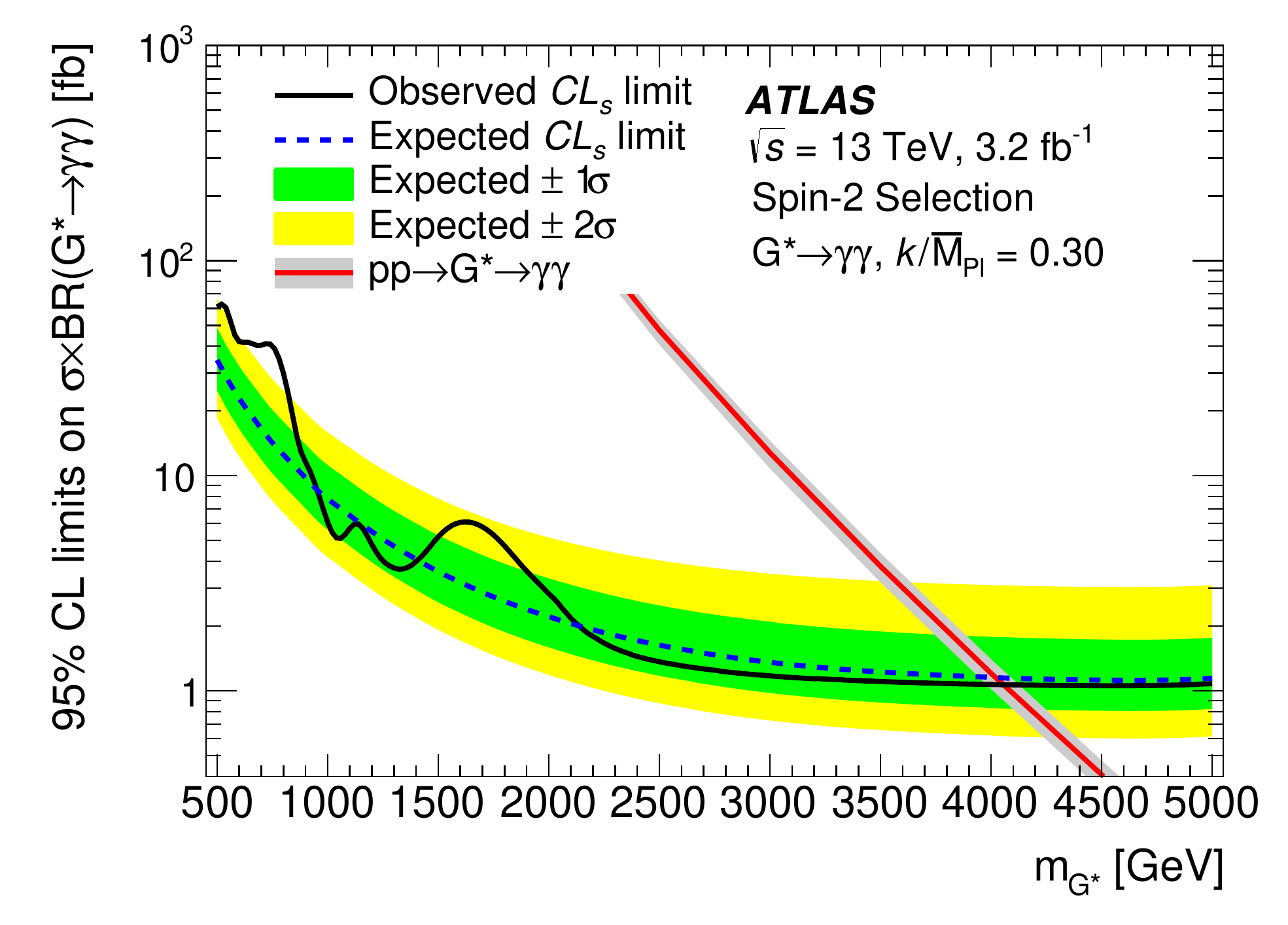}}
\end{center}
\caption{Upper limits on the production cross section of an RS graviton as a function of the assumed mass, for different
values of \coupling\ .
The predicted cross sections  times branching ratio to two photons for the RS graviton model, computed at LO, are also shown. The uncertainty
in the cross-section values represent the PDF uncertainty.
}
\label{fig:spin2_limit}
\end{figure}

Figure~\ref{fig:spin2_limit} shows the corresponding limits on the spin-2 RS graviton cross section times branching ratio
to two photons as a function of mass for different \coupling\
values from 0.01 to 0.3
and the comparison with the cross sections predicted by the benchmark model. The predicted cross sections are computed
at LO in QCD using \textsc{Pythia8}. The uncertainty band represents the PDF uncertainty estimated from the variations of
the NNPDF23LO PDF set.
Outside of the excess region,
the  observed limits on the cross section times branching ratio range from 20~fb to 1~fb for an RS graviton
mass between 500~GeV and 5000~GeV for small couplings and 60~fb to 1~fb for \coupling\ =~0.3.

\begin{figure}[!h]
\begin{center}
\subfloat[]{\includegraphics[width=.5\textwidth]{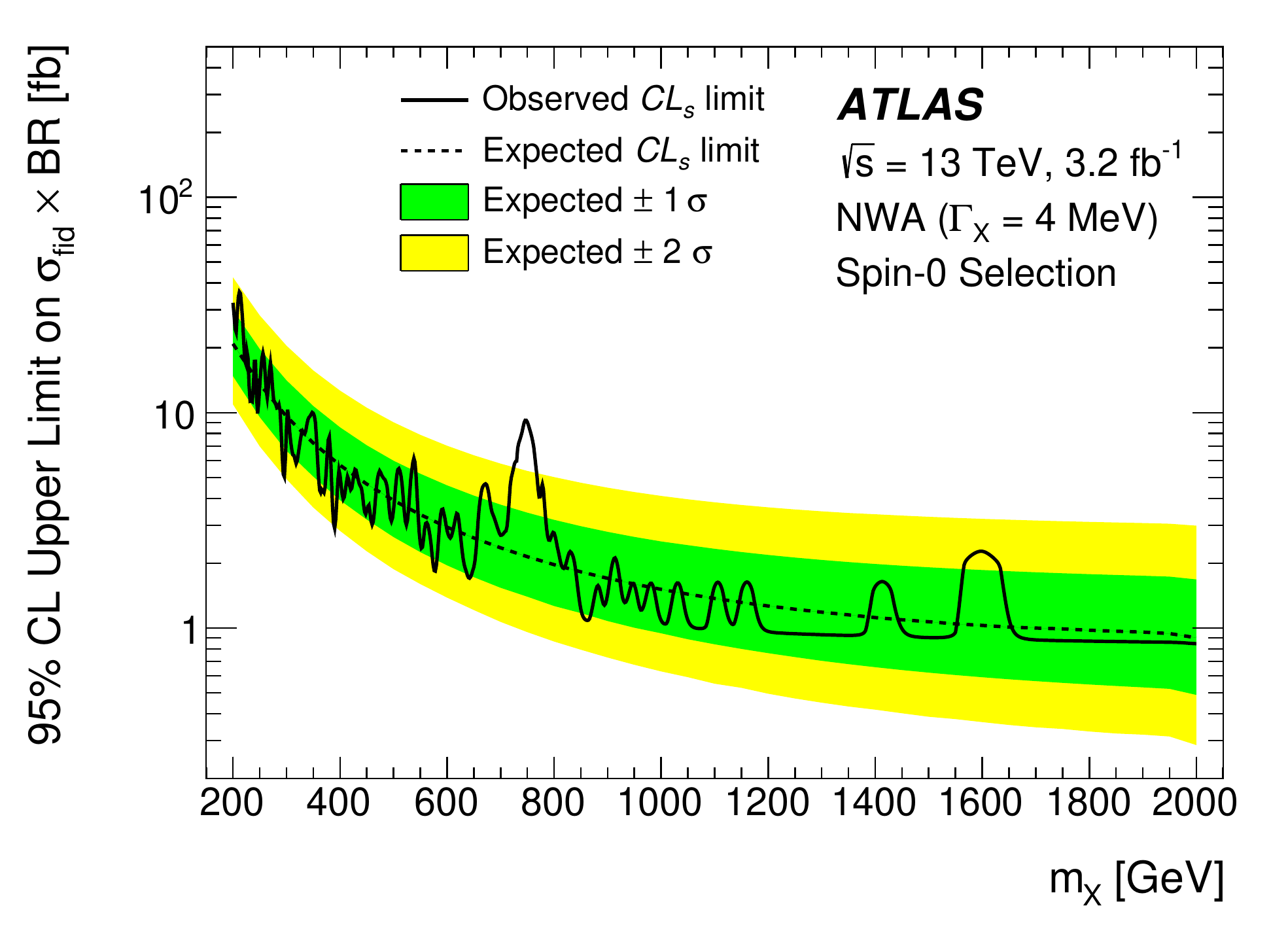}}
\subfloat[]{\includegraphics[width=.5\textwidth]{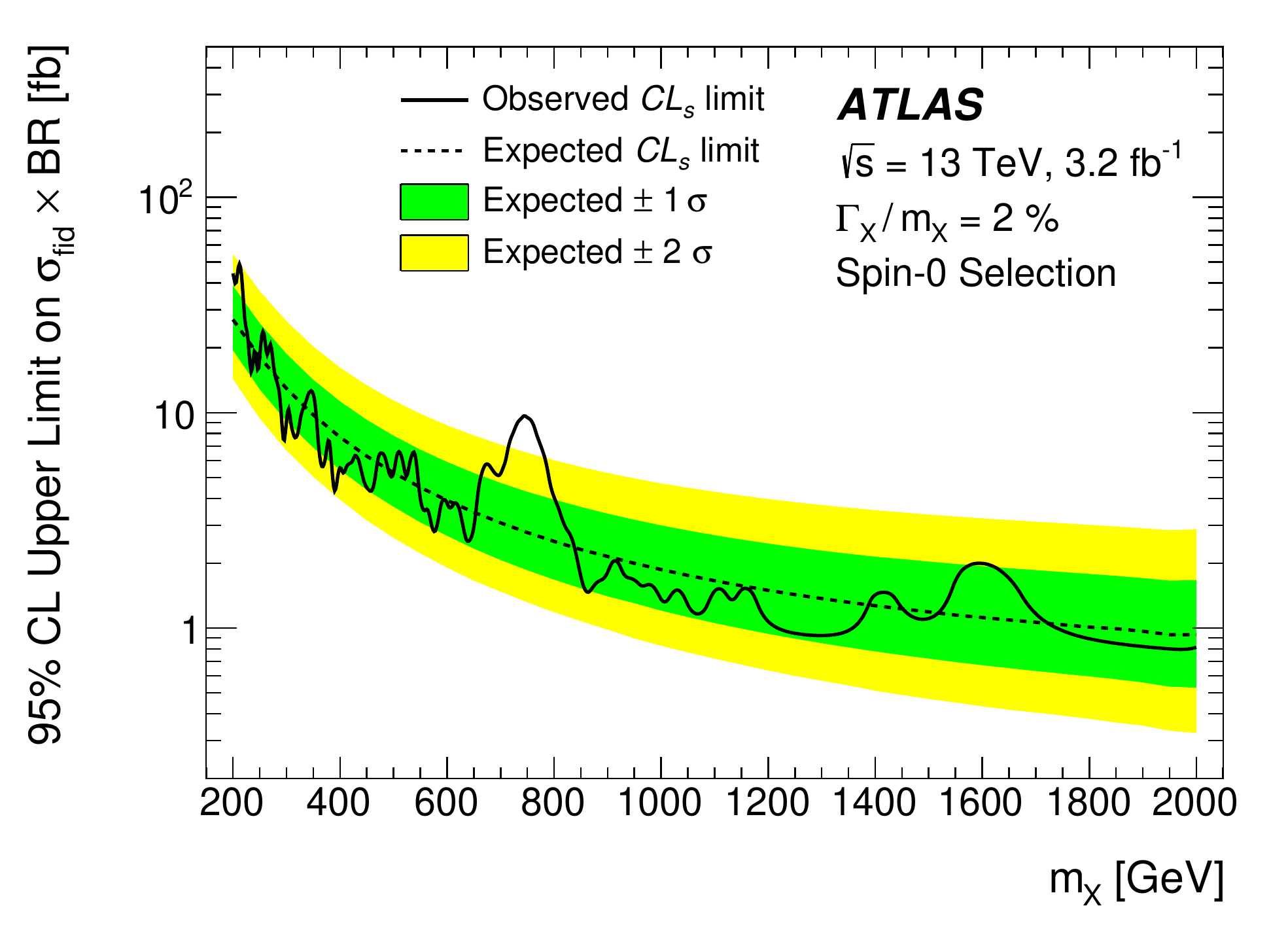}} \\
\subfloat[]{\includegraphics[width=.5\textwidth]{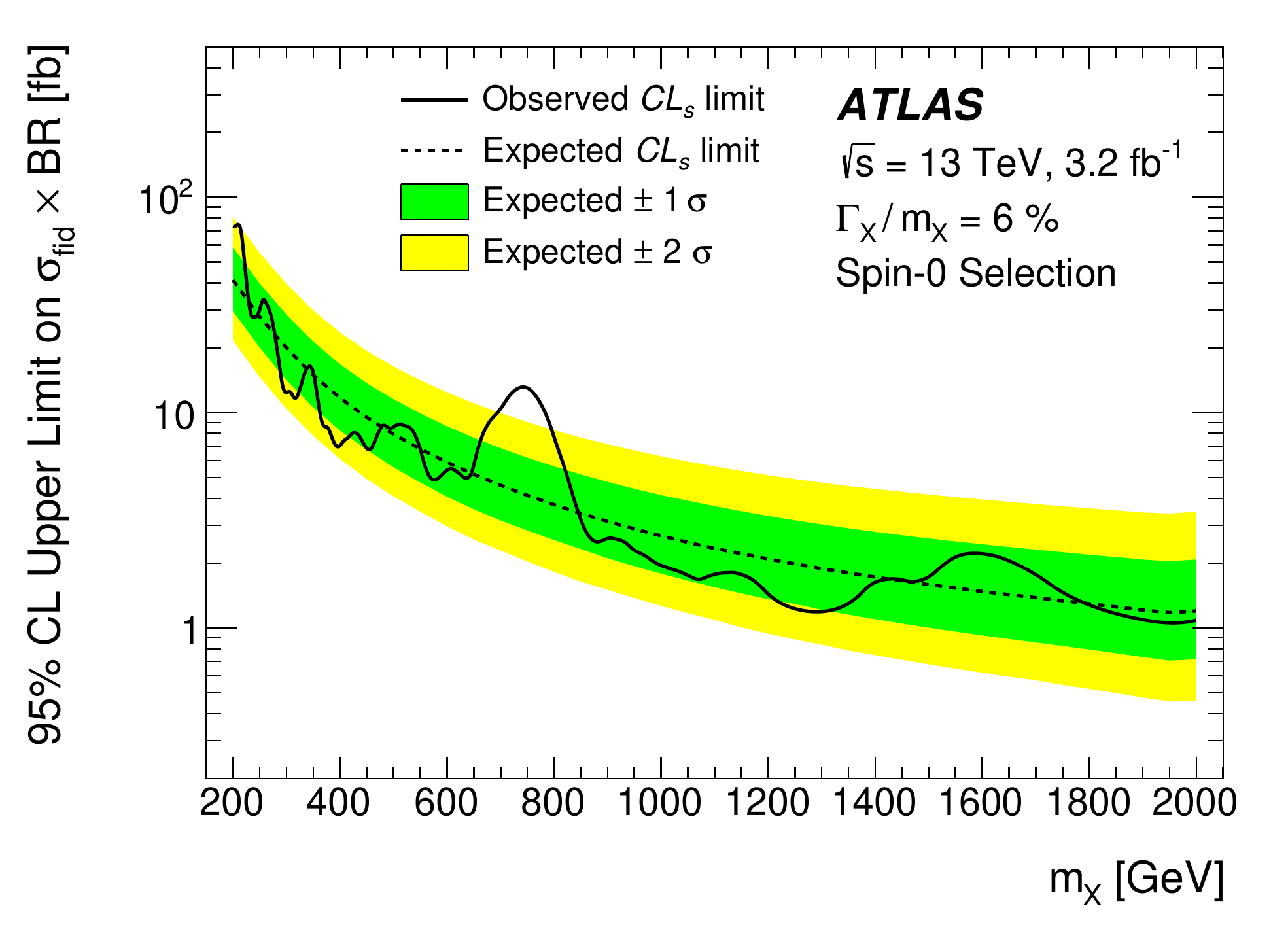}}
\subfloat[]{\includegraphics[width=.5\textwidth]{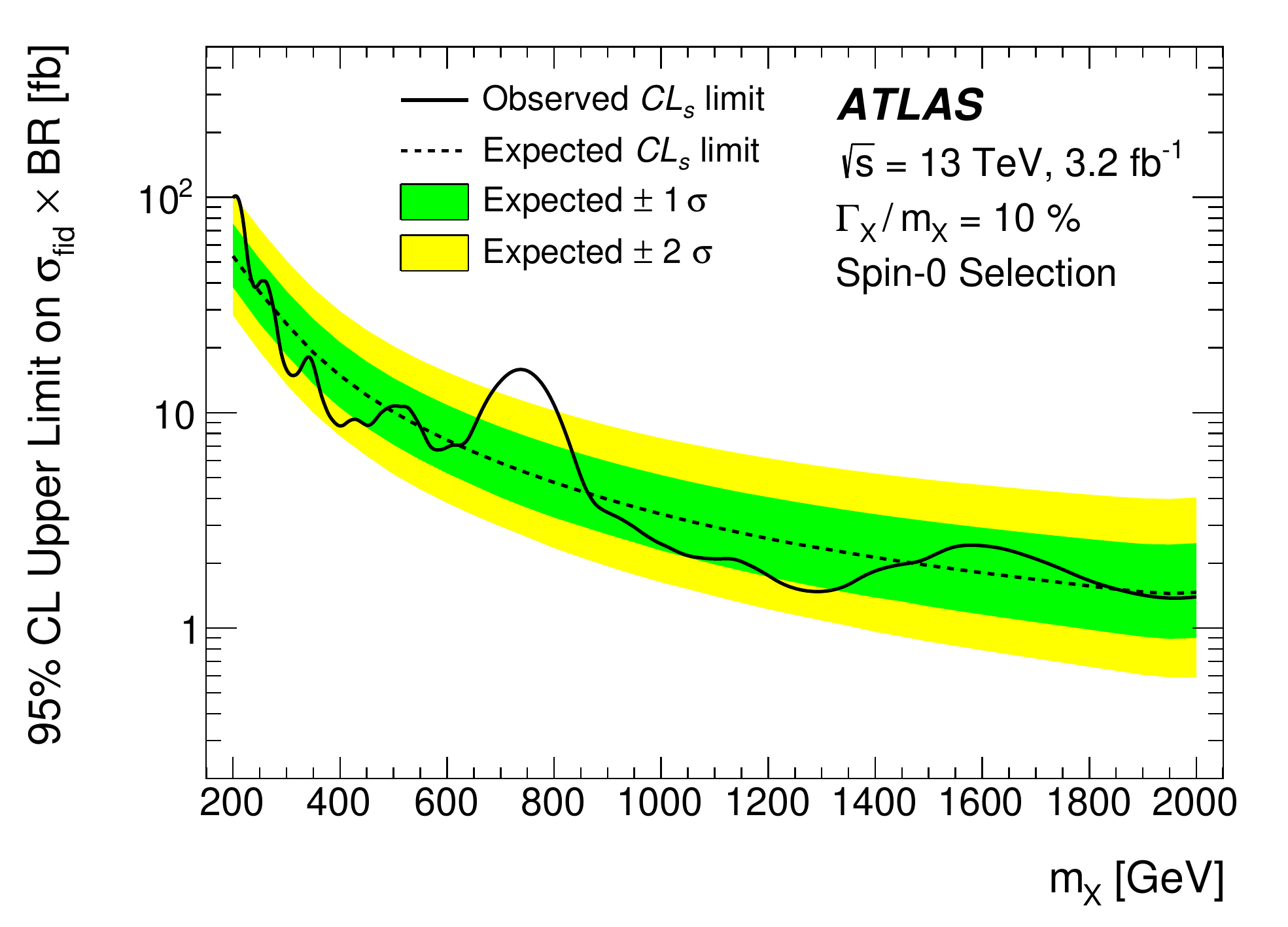}}
\end{center}
\caption{Upper limits on the fiducial cross section at $\sqrt{s} = 13$~TeV of a spin-0 particle as a function of the assumed mass $m_X$, 
for different
values of the decay width divided by the mass. In (a) a narrow-width signal, with $\Gamma$ = 4~MeV, is assumed.}
\label{fig:spin0_limit}
\end{figure}

Figure~\ref{fig:spin0_limit} shows the limits on the signal fiducial cross section times branching ratio to two photons for a spin-0 particle
as a function of the hypothesized mass for various assumptions about the width.
Except near 750~GeV, the observed
limit is in agreement with the expected limit assuming the background-only hypothesis.
For a narrow decay width,  the limits on the fiducial cross section times branching ratio range from
35~fb near 200~GeV to 1~fb at 2000 GeV.

\clearpage

\section{Conclusion}
\label{sec:conclusion}

Searches for new resonances decaying into two photons in the ATLAS experiment at the LHC are presented. 
The $pp$ collision data corresponding to an integrated luminosity of 3.2~fb$^{-1}$ were 
recorded in 2015 at a centre-of-mass energy of $\sqrt{s}$ = 13~TeV. Analyses optimized for the 
search for spin-2 Randall--Sundrum graviton resonances with mass above 500~GeV and for spin-0 resonances with mass
above 200~GeV are performed.
The events selected in the second analysis are a subset of the events selected for the spin-2 search.

Over most of the diphoton mass range, the data are consistent with the background-only hypothesis and 95\% CL 
limits are derived on the cross section for the production of the two benchmark resonances as a function
of their masses and widths. 
Varying both the mass and the decay width of the hypothesised resonance,
the largest deviation from the background-only hypothesis is observed in a broad 
region near a mass of 750~GeV and with a width of about 50~GeV, 
with local significances of 3.8 and 3.9 standard deviations in the searches 
optimized for the spin-2 and spin-0 resonances, respectively. The global significances are estimated to be
2.1 standard deviations for both searches.
When considering narrow-width signal hypotheses, the largest local significances for the two searches are observed near
a mass of 770~GeV and 750~GeV with local significances corresponding to 3.3 and 2.9 standard deviations, respectively.
No significant difference is observed in the properties of the events with a diphoton mass near 750~GeV compared to those
at higher or lower masses.
Assuming a scaling of the production cross section for an $s$-channel resonance produced by
gluon fusion (light quark--antiquark annihilation), the consistency between the
13~TeV data and the data collected at 8~TeV is found to be at the level of 2.7 (3.3) standard deviations
using results from the searches optimized for a spin-2 particle and at the level of
1.2 (2.1) standard deviations using results from the searches optimized for a spin-0 particle.
\section*{Acknowledgements}


We thank CERN for the very successful operation of the LHC, as well as the
support staff from our institutions without whom ATLAS could not be
operated efficiently.

We acknowledge the support of ANPCyT, Argentina; YerPhI, Armenia; ARC, Australia; BMWFW and FWF, Austria; ANAS, Azerbaijan; SSTC, Belarus; CNPq and FAPESP, Brazil; NSERC, NRC and CFI, Canada; CERN; CONICYT, Chile; CAS, MOST and NSFC, China; COLCIENCIAS, Colombia; MSMT CR, MPO CR and VSC CR, Czech Republic; DNRF and DNSRC, Denmark; IN2P3-CNRS, CEA-DSM/IRFU, France; GNSF, Georgia; BMBF, HGF, and MPG, Germany; GSRT, Greece; RGC, Hong Kong SAR, China; ISF, I-CORE and Benoziyo Center, Israel; INFN, Italy; MEXT and JSPS, Japan; CNRST, Morocco; FOM and NWO, Netherlands; RCN, Norway; MNiSW and NCN, Poland; FCT, Portugal; MNE/IFA, Romania; MES of Russia and NRC KI, Russian Federation; JINR; MESTD, Serbia; MSSR, Slovakia; ARRS and MIZ\v{S}, Slovenia; DST/NRF, South Africa; MINECO, Spain; SRC and Wallenberg Foundation, Sweden; SERI, SNSF and Cantons of Bern and Geneva, Switzerland; MOST, Taiwan; TAEK, Turkey; STFC, United Kingdom; DOE and NSF, United States of America. In addition, individual groups and members have received support from BCKDF, the Canada Council, CANARIE, CRC, Compute Canada, FQRNT, and the Ontario Innovation Trust, Canada; EPLANET, ERC, FP7, Horizon 2020 and Marie Sk{\l}odowska-Curie Actions, European Union; Investissements d'Avenir Labex and Idex, ANR, R{\'e}gion Auvergne and Fondation Partager le Savoir, France; DFG and AvH Foundation, Germany; Herakleitos, Thales and Aristeia programmes co-financed by EU-ESF and the Greek NSRF; BSF, GIF and Minerva, Israel; BRF, Norway; Generalitat de Catalunya, Generalitat Valenciana, Spain; the Royal Society and Leverhulme Trust, United Kingdom.

The crucial computing support from all WLCG partners is acknowledged gratefully, in particular from CERN, the ATLAS Tier-1 facilities at TRIUMF (Canada), NDGF (Denmark, Norway, Sweden), CC-IN2P3 (France), KIT/GridKA (Germany), INFN-CNAF (Italy), NL-T1 (Netherlands), PIC (Spain), ASGC (Taiwan), RAL (UK) and BNL (USA), the Tier-2 facilities worldwide and large non-WLCG resource providers. Major contributors of computing resources are listed in Ref.~\cite{ATL-GEN-PUB-2016-002}.

\clearpage

\printbibliography

\newpage

\begin{flushleft}
{\Large The ATLAS Collaboration}

\bigskip

M.~Aaboud$^\textrm{\scriptsize 135d}$,
G.~Aad$^\textrm{\scriptsize 86}$,
B.~Abbott$^\textrm{\scriptsize 113}$,
J.~Abdallah$^\textrm{\scriptsize 64}$,
O.~Abdinov$^\textrm{\scriptsize 12}$,
B.~Abeloos$^\textrm{\scriptsize 117}$,
R.~Aben$^\textrm{\scriptsize 107}$,
O.S.~AbouZeid$^\textrm{\scriptsize 137}$,
N.L.~Abraham$^\textrm{\scriptsize 149}$,
H.~Abramowicz$^\textrm{\scriptsize 153}$
% ...
\bigskip
\\
% ...
$^{12}$ Institute of Physics, Azerbaijan Academy of Sciences, Baku, Azerbaijan\\
% ...
$^{64}$ University of Iowa, Iowa City IA, United States of America\\
% ...
$^{86}$ CPPM, Aix-Marseille Universit{\'e} and CNRS/IN2P3, Marseille, France\\
% ...
$^{107}$ Nikhef National Institute for Subatomic Physics and University of Amsterdam, Amsterdam, Netherlands\\
% ...
$^{113}$ Homer L. Dodge Department of Physics and Astronomy, University of Oklahoma, Norman OK, United States of America\\
% ...
$^{117}$ LAL, Univ. Paris-Sud, CNRS/IN2P3, Universit{\'e} Paris-Saclay, Orsay, France\\
% ...
$^{135}$ $^{(a)}$ Facult{\'e} des Sciences Ain Chock, R{\'e}seau Universitaire de Physique des Hautes Energies - Universit{\'e} Hassan II, Casablanca; $^{(b)}$ Centre National de l'Energie des Sciences Techniques Nucleaires, Rabat; $^{(c)}$ Facult{\'e} des Sciences Semlalia, Universit{\'e} Cadi Ayyad, LPHEA-Marrakech; $^{(d)}$ Facult{\'e} des Sciences, Universit{\'e} Mohamed Premier and LPTPM, Oujda; $^{(e)}$ Facult{\'e} des sciences, Universit{\'e} Mohammed V, Rabat, Morocco\\
% ...
$^{137}$ Santa Cruz Institute for Particle Physics, University of California Santa Cruz, Santa Cruz CA, United States of America\\
% ...
$^{149}$ Department of Physics and Astronomy, University of Sussex, Brighton, United Kingdom\\
% ...
$^{153}$ Raymond and Beverly Sackler School of Physics and Astronomy, Tel Aviv University, Tel Aviv, Israel\\
% ...
$^{a}$ Also at Department of Physics, King's College London, London, United Kingdom\\
$^{b}$ Also at Institute of Physics, Azerbaijan Academy of Sciences, Baku, Azerbaijan\\
$^{c}$ Also at Novosibirsk State University, Novosibirsk, Russia\\
$^{d}$ Also at TRIUMF, Vancouver BC, Canada\\
$^{e}$ Also at Department of Physics {\&} Astronomy, University of Louisville, Louisville, KY, United States of America\\
% ...
\end{flushleft}


\end{document}